\documentclass{sig-alternate-10pt}
\usepackage{graphicx}
\usepackage{amssymb}
\usepackage{epstopdf}
\usepackage{algorithm2e}
\usepackage{cite}
\usepackage[T1]{fontenc}
\usepackage[latin9]{inputenc}
\usepackage{units}
\usepackage{subfig}
\def\E{\mathbb{E}}

\title{Network Coded TCP (CTCP)}
\author{MinJi Kim\thanks{M. Kim, J. Cloud, A. ParandehGheibi, L. Urbina, K. Fouli, and M. M\'edard are with the Massachusetts Institute of Technology, MA USA (e-mail: \{minjikim, jcloud, parandeh, lurbina, fouli, medard\}@mit.edu).} , Jason Cloud$^*$, Ali ParandehGheibi$^*$, Leonardo Urbina$^*$, \\Kerim Fouli$^*$, Douglas Leith\thanks{D. Leith is with the Hamilton Institute, NUI Maynooth, Ireland (e-mail: doug.leith@nuim.ie).} , Muriel M\'edard$^*$}

\begin{document}
\maketitle

\begin{abstract}
We introduce CTCP, a reliable transport protocol using network coding. CTCP is designed to incorporate TCP features such as congestion control, reliability, and fairness while significantly improving on TCP's performance in lossy, interference-limited and/or dynamic networks. A key advantage of adopting a transport layer over a link layer approach is that it provides backward compatibility with wireless equipment installed throughout existing networks.   We present a portable userspace implementation of CTCP and extensively evaluate its performance in both testbed and production wireless networks.
\end{abstract}

%%%%%%%%%%%%%%%%%%%%%%%%
\section{Introduction}

\begin{figure}
\centering
\includegraphics[width=0.75\columnwidth]{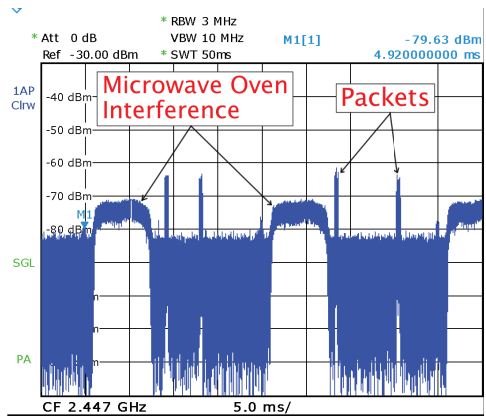}
\caption{Spectrum analyzer measurements showing  microwave oven pulsed interference and 802.11 packet transmissions. The y-axis grid is in 10ms increments. The resolution bandwidth is set to 20 MHz, and thus captures about 99\% of the WLAN signal. }\label{fig:mwo1}
\vspace*{-.3cm}
\end{figure}

Interference, especially for wireless channels located in unlicensed and white-space bands, is a major contributor to packet erasures and can be the primary factor affecting overall network performance. Common sources of interference include not only hidden terminals, but also other wireless links using heterogeneous technologies or non-network wireless devices such as radar, microwave ovens, etc.  As an example, Figure \ref{fig:mwo1} shows microwave oven interference that leads to packet erasures in 802.11 networks operating within the same or adjacent channels. With this type of interference, many existing transport layer protocols, such as TCP, are known to perform poorly and a variety of cross-layer techniques have been developed to increase performance \cite{bakshi97}.  Furthermore, a large majority of research has primarily focussed on a range of link layer approaches for mitigating the affects of such interference. In this paper we adopt a different strategy and propose a novel transport layer solution referred to as Network Coded TCP (CTCP).

The adoption of a transport layer approach ensures backward compatibility with legacy equipment. A recent industry study \cite{technicolor} estimates that almost 1.2 billion 802.11 devices have been shipped to date.  Replacing these devices, in addition to other existing wireless infrastructures, in order to incorporate new link layer technology is largely impractical due to the costs involved.  Furthermore, the need to provide a portable, easily implemented solution is also required to overcome challenges due to the proliferation of operating systems in use throughout the internet. This strongly motivates techniques that have the capability to be retro-fitted into a wide variety of existing systems with the aim of bringing immediate benefits to current and future users.  We provide a userspace implementation so that adoption of the techniques presented in this paper can be achieved regardless of the user's operating system or network infrastructure.

We further demonstrate how the TCP transport layer can be combined with network coding, without the need to consider cross-layer techniques such as explicit feedback from the link layer or other techniques such as explicit congestion notification (ECN), to yield significant performance gains in the presence of interference. For example, our measurements in public WiFi hotspots around Boston (see later) demonstrate that CTCP can achieve reductions in connection completion times of 100-300\% compared with uncoded TCP.   In controlled lab experiments, we consistently observed reductions of more than an order of magnitude (i.e. >1000\%) in completion times for both HTTP and streaming video flows when the link packet loss rate exceeds 5\%. By using a transport layer solution, we can provide error-correction coding \emph{across} packets rather than simply within each individual packet so that CTCP yields significant performance gains in many complex, time-varying environments.

Our contributions include the following.  First, we develop an error-correction coding approach tailored to packet streams with delayed feedback from the receiver.   Classical error-correction coding for packet erasure channels is largely either open-loop in nature (i.e. forward error correction) or assumes near instantaneous feedback from the receiver (akin to automatic repeat request (ARQ)).    We develop a variable-length block coding scheme that makes efficient use of delayed feedback to achieve high throughput and low decoding delay.   We demonstrate that this scheme can be practically implemented in software on commodity hardware.   Second, standard TCP congestion control uses packet loss as an indicator of congestion. The masking of path losses by error correction coding necessitates the development of a novel TCP congestion control approach.   In keeping with the overall approach taken here, we seek to make the smallest possible changes consistent with achieving high performance.  We propose a simple change to the TCP additive-increase, multiplicative-decrease (AIMD) algorithm.  This algorithm introduces an adaption of the backoff factor, which reverts to the standard AIMD operation in networks where packets losses primarily occur due to queue overflow (e.g. most wired networks) ensuring backward compatibility with other congestion control algorithms.   Third, we carry out an extensive experimental evaluation of the CTCP system both in controlled lab conditions (enabling reproducible, detailed performance analysis) and in production wireless networks (providing network conditions and performance data in the ``wild'').

%%%%%%%%%%%%%%%%%%%%%%%%
\section{Related Work}\label{sec:background}

Perhaps the closest previous work is that of Sundararajan et al.\cite{tcpnc}, which introduced TCP/NC and proposes a sliding window coding scheme that is fundamentally different from the variable-length block coding scheme proposed here.  Their work focuses on a theoretical analysis over idealized links, and does not propose a method to guarantee intermediate packet decoding at the receiver.   The use of coding has also been proposed before our work or that of \cite{tcpnc} in order to mitigate the effect of losses on TCP, particularly in the context of wireless links.  We may roughly taxonomize these into approaches that consider the operation of TCP jointly with  lower layer redundancy, at the physical or MAC level, often in a cross-layer approach \cite{bakshi97, BF03, BMAA04, BA02, LGT02, CZT99, KBG10, MSM02, AHR08, TFHSM07} and those that consider redundancy at the IP layer or above without access to the lower layers \cite{PP10, TSKK05, BLK04, ACV04, LK04, PCM00, Sub08}. Since our goal is to provide a system that operates without access to, or even detailed knowledge of, the physical layer, it is in the latter category that our approach belongs.  

The approaches in papers implementing coding at the IP layer for operation with TCP have generally revolved around traditional fixed-length block codes, such as Reed-Solomon codes. Furthermore, many of these approaches are combined with mechanisms such as ECN \cite{TSKK05, MSM02} to aid in generating the necessary redundancy. These block-based approaches acknowledge packets upon successful decoding and, if the number of errors exceeds the predetermined level of redundancy, a decoding failure occurs, requiring retransmission of the full block. Our approach uses network coding to mitigate this problem.  If the number of errors exceeds the provided level of redundancy, additional redundancy can be provided by sending additional degrees of freedom (dof), or coded packets, rather than retransmitting the entire block.

%%%%%%%%%%%%%%%%%%%%%%%%
\section{Overview of Coded TCP (CTCP)}\label{sec:overview}

Before discussing the CTCP sender and receiver in detail, we provide a more holistic view of the two here. The CTCP sender initially segments the stream, or the file, into a series of blocks containing $blksize$ number of packets, where each packet is assumed to be of fixed length. If the remainder of a file or a stream is not large enough to form a complete packet, the packet is padded with zeros to ensure that all packets are of the same length. A block need not be completely full, i.e. a block may have fewer than $blksize$ packets; however, block $i$ should be full before block $i+1$ is initialized.   After transmitting the initial block of packets, the size of the block is then adapted in light of feedback from the receiver.

The CTCP sender buffers $numblks$ of blocks and the value of $numblks$ should be conveyed to the receiver. The value of $numblks$ may be negotiated at initialization between the sender and the receiver, as $numblks$ directly affects the memory usage on both ends. We denote the smallest block in memory to be $currblk$. Note that this does not mean that CTCP sender may send $numblks\times blksize$ amount of data at any time.

The sender is allowed to transmit packets only if the congestion control mechanism allows it to; however, whenever it is allowed to transmit, the sender may choose to transmit a packet from any one of the blocks in memory, i.e. blocks $currblk$, $currblk+1$, ..., $currblk+numblks-1$. In Section \ref{sec:blockschedule}, we shall discuss the sender's algorithm for selecting a block from which it sends a packet. The payload of the transmitted packet may be coded or uncoded; the details of the coding operations can be found in Section \ref{sec:coding}.

The sender includes the following in each packet: (i) the block number, (ii) a seed for a pseudo-random number generator which allows the receiver to generate the coding coefficients, (iii) the sequence number $seqno$, and (iv) the (coded or uncoded) payload.  The sequence number for CTCP differs from that of TCP -- for TCP, a sequence number indicates a specific data byte; for CTCP, a sequence number indicates that a packet is the $seqno$-th packet transmitted by the sender, thus, is not tied to a byte in the file.

The CTCP receiver sends acknowledgments (ACKs) for the packets it receives. In the ACK, the receiver indicates: (i) the smallest undecoded block $ack\_currblk$, (ii)  the number of degrees of freedom (dofs) $ack\_currdof$ it has received for the current block $ack\_currblk$, and (iii)  the $ack\_seqno$ of the packet it is acknowledging.

Using the information in an ACK, the CTCP sender adjusts its behavior. We first describe the sender algorithm in Section \ref{sec:sender}, as most of the intelligence of the protocol is on the sender's side. Then, we present the CTCP receiver algorithm in Section \ref{sec:receiver}. The CTCP receiver's main role is to decode and deliver data to the application.

%%%%%%%%%%%%%%%%%%%%%%%%
\section{CTCP Sender}\label{sec:sender}

We present the sender side algorithm for CTCP. The sender maintains several internal parameters (defined in Table \ref{tab:defintions}), which it uses to generate coded packets and schedule blocks to be transmitted.

\begin{table}[tb]
\small
\caption{Definitions of the sender parameters}\label{tab:defintions}
\begin{tabular}{|l|p{5.5cm}|}
\hline
Notation & Definition\\
\hline
\hline
$p$ & Short term average packet loss rate\\
$RTO$ & Retransmission timeout period (equal to $\gamma\cdot RTT$ where $\gamma\geq 1$ is a constant)\\
$RTT$ & Current (or the last acknowledged packet's) round-trip time\\
$RTT_{min}$ & The minimum round-trip time\\
$seqno\_nxt$ & The sequence number of the next packet to be transmitted\\
$seqno\_una$ & The sequence number of the latest unacknowledged packet\\
$ss\_threshold$ & Slow-start threshold, i.e. if $tokens > ss\_threshold$, the sender leaves the slow-start mode\\
$time\_lastack$ & Timestamp of when the sender received the latest ACK (initialized to the time when the sender receives a SYN packet from the receiver)\\
$tokens$ & Number of tokens, which is conceptually similar to congestion window for traditional TCP\\
$blksize$ & Size of the blocks (in number of packets)\\
$currblk$ & Current block number at the sender, which is the smallest unacknowledged block number\\
$currdof$ & Number of dofs the receiver has acknowledged for the current block\\
$numblks$ & Number of active blocks, i.e. the sender may schedule and transmit packets from blocks $currblk$, $currblk+1$, ..., $currblk+numblks-1$\\
$B(seqno)$ & Block number from which packet with $seqno$ was generated from\\
$T(seqno)$ & Timestamp of when packet with $seqno$ was sent\\
\hline
\end{tabular}
\vspace*{-.3cm}
\end{table}

%%%%%%%%%%%%%%%%%%%%%%%%
\subsection{Network Parameter Estimation}\label{sec:estimation}

The CTCP sender estimates the network parameters, such as $RTT$ and estimated path loss rate $p$, using the received ACKs as shown in Algorithm \ref{alg:update}. The sender adjusts its actions, including coding operations and congestion control depending on these estimates. If the sender does not receive an ACK from the receiver for an extended period of time (i.e. a time-out occurs), the network parameters are reset to predefined default values. The predefined default values may need to be chosen with some care such that they estimate roughly what the network may look like.

\begin{algorithm}[tb]
\small
Receive an ACK\;
$time\_lastack \leftarrow$ current time\;
$RTT \leftarrow time\_lastack - T(ack\_seqno)$\;
$RTT_{min} \leftarrow$ minimum of $\{RTT, RTT_{min}\}$\;
\If{$ack\_currblk > currblk$}
{
Free blocks $currblk$, ..., $ack\_currblk-1$\;
$currblk \leftarrow ack\_currblk$\;
$currdof \leftarrow ack\_currdof$\;
}
$currdof \leftarrow \max\{ack\_currdof, currdof\}$\;
\If{$ack\_seqno > seqno\_una$}
{
$losses \leftarrow ack\_seqno - seqno\_una + 1$\;
$p \leftarrow p(1-\mu)^{losses+1}+(1-(1-\mu)^{losses})$\;
}
$seqno\_una \leftarrow ack\_seqno+1$\;
\vspace*{.2cm}
\caption{CTCP sender algorithm for updating the network parameters.
\vspace{-0.5cm}}\label{alg:update}
\end{algorithm}

The CTCP sender maintains moving averages of $p$. We use a slightly modified version of the exponential smoothing technique, where we consider the data series we are averaging to be a 0-1 sequence, where 0 indicates that the packet has been received successfully and 1 otherwise (i.e. a packet loss). Now, assume that there were $losses$ number of packets lost. If $losses = 0$, then the update equation for $p$ in Algorithm \ref{alg:update} becomes
\begin{equation}
p \leftarrow p(1-\mu)+0,
\end{equation}
where $\mu$ is the smoothing factor. If $losses = 1$, the same update equation becomes
\begin{equation}
p\leftarrow p(1-\mu)^{2}+\mu= (1-\mu)[p (1-\mu) + 0 ] + \mu,
\end{equation}
which is identical to executing an exponential smoothing over two data points (one lost and one acknowledged). We can repeat this idea for $losses > 1$ to obtain the update rule for $p$ in Algorithm \ref{alg:update}. Therefore, the fact that a single ACK may represent multiple losses ($losses\geq 1$) leads to a slightly more complicated update rule for $p$ than that for $RTT$ as shown in Algorithm \ref{alg:update}. To the best of our knowledge, such an update rule has not been used previously.

%%%%%%%%%%%%%%%%%%%%%%%%
\subsection{Reliability}\label{sec:reliability}

CTCP achieves reliability by using receiver feedback to dynamically adapt the size of each block such that it can be successfully decoded. In Algorithm \ref{alg:update}, the CTCP sender increments $currblk$ only if it has received an ACK indicating that the receiver is able to decode $currblk$ (i.e. $ack\_currblk > currblk$). This mechanism is similar to traditional TCP's window sliding scheme in which the TCP sender only slides its window when it receives an ACK indicating the some bytes have been received. In the case of CTCP, the reliability is implemented over blocks instead of bytes.

%%%%%%%%%%%%%%%%%%%%%%%%
\subsection{Congestion Control Mechanism}\label{sec:congestion}

Traditional TCP's AIMD congestion control increases the TCP sender's congestion window size $cwnd$ by $\alpha$ packets per RTT and multiplicatively decreases $cwnd$ by a backoff factor $\beta$ on detecting packet losses within one RTT inducing a single $cwnd$ backoff. The usual values are $\alpha=1$ when appropriate byte counting is used, and $\beta=0.5$.   On lossy links, repeated backoffs in response to noise losses rather than queue overflow can prevent $cwnd$ from increasing to fill the available link capacity.   The behavior is well known and is captured, for example, in %the Padhye model
\cite{padhye} in which $cwnd$ scales as $\sqrt{1.5/p}$, where $p$ is the packet loss rate.

The basic issue here is that on lossy links, loss is not a reliable indicator of network congestion.  One option might be to use delay, rather than loss, as the indicator of congestion, but this raises many new issues and purely delay-based congestion control approaches have not been widely adopted in the internet despite being the subject of extensive study.    Another option might be to use explicit signalling, for example via ECN, but this requires both network-wide changes and disabling of $cwnd$ backoff on packet loss.   These considerations motivate consideration of hybrid approaches, making use of both loss and delay information.   The use of hybrid approaches is well-established, for example Compound TCP \cite{compoundtcp} is widely deployed.

We consider modifying the AIMD multiplicative backoff. Before discussing the details of backoff behavior, we emphasize that CTCP uses \emph{tokens} to control the CTCP sender's transmission rate instead of the congestion window $cwnd$. Therefore, $tokens$ play a similar role for CTCP as $cwnd$ does for TCP. A token allows the CTCP sender to transmit a packet (coded or uncoded). When the sender transmits a packet, the token is used. The number of tokens, $tokens$, is controlled according to the modified AIMD multiplicative backoff (Algorithm \ref{alg:congestion}).

As shown in Algorithm \ref{alg:congestion}, we modify the AIMD multiplicative backoff to have
\begin{align}\label{eq:beta}
\beta = \frac{RTT_{min}}{RTT},
\end{align}
where $RTT_{min}$ is the path round-trip propagation delay (which is typically estimated as the lowest per packet RTT observed during the lifetime of a connection) and $RTT$ is the current round-trip time.

\begin{algorithm}[tbp]
\small
\If{\emph{current time} $> time\_lastack + RTO$}
{
$tokens \leftarrow$ initial token number\;
Set to slow-start mode\;
}
\If{\emph{Receive an ACK on path} $i$}{
\eIf{\emph{slow-start mode}}
{
$tokens \leftarrow tokens + 1$\;
\If{$tokens > ss\_threshold$}
{
Set to congestion avoidance mode\;
}
}
{
\eIf{$ack\_seqno > seqno\_una$}
{
$tokens \leftarrow \frac{RTT_{min}}{RTT}tokens $\;
}
{
$tokens \leftarrow tokens + \frac{1}{tokens}$\;
}
}

}
\vspace*{.2cm}
\caption{CTCP sender congestion control.}\label{alg:congestion}
\end{algorithm}

This is similar to the approach considered in \cite{backoff}, which uses $\beta = RTT_{min}/RTT_{max}$ with the aim of making TCP throughput performance less sensitive to the level of queue provisioning.    Indeed on links with only queue overflow losses, (\ref{eq:beta}) reduces to the approach in \cite{backoff} since $RTT=RTT_{max}$ (the link queue is full) when loss occurs.   In this case, when a link is provisioned with a bandwidth-delay product of buffering, as per standard guidelines, then $RTT_{max}=2RTT_{min}$ and $\beta=0.5$, i.e. the behavior is identical to that of standard TCP. More generally, when queue overflow occurs the sum of the flows' throughputs must equal the link capacity $B$, $\sum_{i=1}^n tokens_i/RTT_i=B$ where $n$ is the number of flows.  After backoff according to (\ref{eq:beta}), the sum-throughput becomes $\sum_{i=1}^n \beta_i tokens_i/RTT_{min,i} = B$ when the queue empties.   That is, the choice (\ref{eq:beta}) for $\beta$ decreases the flow's $tokens$ so that the link queue just empties and full throughput is maintained.

On lossy links (with losses in addition to queue overflow losses),  use of $RTT$ in (\ref{eq:beta}) adapts $\beta$ to each loss event.   When a network path is under-utilized, $RTT=RTT_{min}$ (therefore, $\beta=1$ and $\beta\times tokens = tokens$). Thus, $tokens$ is not decreased on packet loss.  Hence, $tokens$ is able to grow, despite the presence of packet loss.   Once the link starts to experience queueing delay,  $RTT>RTT_{min}$ and $\beta<1$, i.e. $tokens$ is decreased on loss.    Since the link queue is filling, the sum-throughput before loss is $\sum_{i=1}^n tokens_i/RTT_i=B$.    After decreasing $tokens$, the sum-throughput is at least  (when all flows backoff their $tokens$) $\sum_{i=1}^n \beta_i tokens_i/RTT_{min,i} = B$ when the queue empties.  That is, (\ref{eq:beta}) adapts $\beta$ to maintain full throughput.

Although we focus on using (\ref{eq:beta}) in combination with linear additive increase (where $\alpha$ is constant), we note that this adaptive backoff approach can also be combined with other types of additive increase including, in particular, those used in Cubic TCP and Compound TCP.  As shown here, these existing approaches can be extended to improve performance on lossy links.

%%%%%%%%%%%%%%%%%%%%%%%%
\subsubsection{Mathematical Modeling}
In this section, we provide some analysis on our choice of $\beta$, the multiplicative backoff factor. Consider a link shared by $n$ flows.    Let $B$ denote the capacity of the link and $T_i$ the round-trip propagation delay of flow $i$.  We will assume that the queueing delay can be neglected, i.e. the queues are small or the link is sufficiently lossy that the queue does not greatly fill.   We also assume that any differences in the times when flows detect packet loss (due to the differences in path propagation delay) can be neglected.   Let $t_k$ denote the time of the $k$-th network backoff event, where a network backoff event is defined to occur when one or more flows reduce their $tokens$.   Let $w_i(k)$ denote the $tokens$ of flow $i$ immediately before the $k$-th network backoff event and $s_i(k)=w_i(k)/T_i$ the corresponding throughput.   With AIMD, we have
\begin{equation}\label{eq:aimd}
s_i(k) = \tilde{\beta}_i(k-1)s_i(k-1) + \tilde{\alpha}_i T(k),
\end{equation}
where $\tilde{\alpha}_i=\alpha/T_i^2$, $\alpha$ is the AIMD increase rate in packets per RTT, $T(k)$ is the time in seconds between the $k-1$ and $k$-th backoff events, and $\tilde{\beta}_i(k)$ is the backoff factor of flow $i$ at event $k$.    The backoff factor $\tilde{\beta}_i(k)$  is a random variable, which takes the value $1$ when flow $i$ does not experience a loss at network event $k$, and takes the value given by (\ref{eq:beta}) otherwise.     The time $T(k)$ is also a random variable, with distribution determined by the packet loss process and typically coupled to the flow rates $s_i(k)$, $i=1,\cdots,n$.

For example,  associate a random variable $\delta_j$ with packet $j$, where $\delta_j=1$ when packet $j$ is erased and $0$ otherwise.  Assume the $\delta_j$ are i.i.d with erasure probability $p$.    Then $Prob(T(k)\le t)=1-(1-p)^{N_t(k)}$ where $N_t(k)=\sum_{i=1}^n N_{f,i}(t)$ is the total number of packets transmitted over the link in interval $t$ following backoff event $k-1$ and $N_{t,i}(k)=\tilde{\beta}_i(k-1)s_i(k-1)t + 0.5\tilde{\alpha}_i t^2$ is the number of packets transmitted by flow $i$ in this interval $t$.   Also, the probability $\gamma_i(k):=Prob(\tilde{\beta_i(k)}=1)$ that flow $i$ does not back off at the $k$-th network backoff event is the probability that it does see any loss during the RTT interval $[T(k),T(k)+T_i]$, which can be approximated by $\gamma_i(k) = (1-p)^{s_i(k)T_i}$ on a link with sufficiently many flows.

Since both $\tilde{\beta}_i(k)$ and $T(k)$ are coupled to the flow rates $s_i(k)$, $i=1,\cdots,n$, analysis of the network dynamics is generally challenging.    When the backoff factor $\tilde{\beta}_i(k)$ is stochastically independent of the flow rate $s_i(k)$, the analysis is then relatively straightforward.   Note that this assumption is valid in a number of useful and interesting circumstances.  One such circumstance is when links  are loss-free (with only queue overflow losses) \cite{aimdmodel}.  Another is on links with many flows and i.i.d packet losses, where the contribution of a single flow $i$ to the queue occupancy (and so to $RTT$ in (\ref{eq:beta})) is small. Further, as we will see later, experimental measurements indicate that analysis using the assumption of independence accurately predicts performance over a range of other network conditions, and so results are relatively insensitive to this assumption.

Given independence, from (\ref{eq:aimd}),
\begin{align}\label{eq:aimd2}
\E[s_i(k)] = \E[\tilde{\beta}_i(k)] \E[s_i(k-1)] + \tilde{\alpha}_i \E[T(k)].
\end{align}
When the network is also ergodic, a stationary distribution of flow rates exists. Let $\E[s_i]$ denote the mean stationary rate of flow $i$.  From (\ref{eq:aimd2}) we have
\begin{align}\label{eq:aimd3}
\E[s_i] =  \frac{\tilde{\alpha}_i }{1-\E[\tilde{\beta}_i]}\E[T].
\end{align}
Since the factor $\E[T]$ is common to all flows, the fraction of link capacity obtained by flow $i$ is determined by $\tilde{\alpha}_i/(1-\E[\tilde{\beta}_i])$.

\noindent\emph{\textbf{Fairness between flows with same RTT:}}  From (\ref{eq:aimd3}), when flows $i$, $j$ have the same RTT, and so $\tilde{\alpha}_i=\tilde{\alpha}_j$, and the same mean backoff factor $\E[\tilde{\beta}_i]=\E[\tilde{\beta}_j]$ then they obtain on average the same throughput share.

\noindent\emph{\textbf{Fairness between flows with different RTTs:}}  When flows $i$, $j$ have different round trip times $T_i\ne T_j$ but the same mean backoff factor, the ratio of their throughputs is $\E[s_i]/\E[s_j] =(T_j/T_i)^2$.  Observe that this is identical to standard TCP behavior \cite{aimdmodel}.

\noindent\emph{\textbf{Fairness between flows with different loss rates:}}  The stationary mean backoff factor $\E[\tilde{\beta}_i]$ depends on the probability that flow $i$ experiences a packet loss at a network backoff event.   Hence, if two flows $i$ and $j$ experience different per packet loss rates $p_i$ and $p_j$ (e.g. they might have different access links while sharing a common throughput bottleneck), this will affect fairness through $\E[\tilde{\beta}_i]$.

\noindent\emph{\textbf{Friendliness:}}  The model (\ref{eq:aimd}) is sufficiently general enough to include AIMD with fixed backoff factor, as used by standard TCP.    We consider two cases.  First, when the link is loss-free (the only losses are due to queue overflow) and all flows backoff when the queue fills, then $1-\E[\tilde{\beta}_i] = 1-\beta_i(k)$.   For a flow $i$ with fixed backoff of 0.5 and a flow $j$ with adaptive backoff $\beta_j$, the ratio of the mean flow throughputs is $\E[s_i]/\E[s_j] =2(1-\beta_j)$ by (\ref{eq:aimd3}) when the flows have the same RTT.   When $\beta_j=T_j/RTT_j=0.5$, the throughputs are equal.  Since $RTT_j=T_j+q_{max}/B$ where $q_{max}$ is the link buffer size and $B$ the link rate, $\beta_j=0.5$ when $q_{max}=BT_j$ (i.e., the buffer is size at the bandwidth-delay product).   The second case we consider here is when the link has i.i.d packet losses with probability $p$.   When $p$ is sufficiently large so that the queue rarely fills, queue overflow losses are rare and the throughput of a flow $i$ with fixed backoff of 0.5 is accurately modeled by the Padhye model \cite{padhye}.   That is, the throughput is largely decoupled from the behavior of other flows sharing the link (since coupling takes place via queue overflow) and, in particular, this means that flows using adaptive backoff do not penalize flows which use fixed backoff.   We present experimental measurements confirming this behavior in Section \ref{sec:friendliness}.

%%%%%%%%%%%%%%%%%%%%%%%%
\subsection{Network Coding Operations}\label{sec:coding}

The coding operations are performed over blocks. 
Setting the initial bock size $blksize = 1$ leads to operations similar to that of traditional TCP variants which cannot effectively take advantage of network coding. On the other hand, setting $blksize$ too large leads to increased encoding/decoding complexity and delay. Therefore, $blksize$ has to be chosen with care.  In our experience, it is desirable to set $blksize$ to be similar to the bandwidth$\times$delay of the network, so that feedback from the receiver about the first packet in the block arrives at the sender around the time that sending of $blksize$ packets is completed and the feedback can then be used to adapt the block size on the fly. 
The choice of coding field size also affects performance. While a higher field size leads to a higher probability of generating independent dofs (resulting in increased efficiency), this comes at the cost of coding and decoding complexity.  We find that using a field of $\mathbb{F}_{256}$, i.e. each coefficient is a single byte, provides a good balance.

To avoid encoding delay and reduce encoding complecity, we use a systematic code -- i.e. uncoded packets are first transmitted and then extra coded packets are sent. In generating coded packets, there are many options. 

 In our design, we use a simple approach -- a coded packet is generated by randomly coding all packets in the block. This approach is most effective in terms erasure correction. With high probability, each coded packet will correct for any single erasure in the block.

%%%%%%%%%%%%%%%%%%%%%%%%
\subsection{Transmission  Scheduling}\label{sec:blockschedule}

\begin{algorithm}[tbp]
\small
Initialize an array $onfly[]$ to $0$\;
\For{$seqno$ \emph{in} $[seqno\_una, seqno\_nxt -1 ]$}
{
\If{\emph{current time} $< T(seqno) + 1.5 RTT$}
{
$onfly[B(seqno)] \leftarrow onfly[B(seqno)] + 1$\;
}
}
\For{$blkno$ \emph{in} $[currblk, currblk+numblks-1]$}{
\uIf{$blkno = currblk$ \emph{and} $(1-p)onfly[currblk] <  blksize - currdof$}
{
Transmit a packet with sequence number $seqno\_nxt$ from block $blkno$\;
$seqno\_nxt \leftarrow seqno\_nxt + 1$\;
}\ElseIf{$(1-p)onfly[blkno] < blksize$}{
Transmit a packet with sequence number $seqno\_nxt$ from block $blkno$\;
$seqno\_nxt \leftarrow seqno\_nxt + 1$\;
}
}
\vspace*{.2cm}
\caption{CTCP sender algorithm for block scheduling when a token is available.}\label{alg:block}
\end{algorithm}

When a token is available, a packet is transmitted from a block decided on by the CTCP sender. The block scheduling algorithm is detailed in Algorithm \ref{alg:block}. The algorithm first estimates the number of packets in a block that are in transit from the sender to the receiver. Given $p$, the sender can then compute the expected number of packets the receiver will receive for any given block. In determining the expected number of dofs the receiver will receive for any given block, we exclude the packets that have been transmitted more than $1.5\cdot RTT$ time ago since they are likely to be lost or significantly delayed. The constant factor of 1.5 may be adjusted depending on the delay constraints of the application of interest; however, the constant factor should be $\geq 1$.

The goal of the sender is to ensure that the receiver will receive enough packets to decode the block, while also limiting the number of unnecessary coded packets transmitted. The sender prioritizes block $i$ before $i+1$; therefore, $currblk$ is of the highest priority. Note that the algorithm treats $currblk$ slightly differently from the rest of the blocks. In our design, the CTCP receiver informs the sender of how many dofs it has received ($currdof$) for block $currblk$. Therefore, the sender is able to use the additional information to determine more precisely whether another packet should be sent from block $currblk$ or not. It is not difficult to piggy-back more information on the ACKs. For example, we could include how many dofs the receiver has received for blocks $currblk$ as well as $currblk+1$, $currblk+2$, ..., $currblk+numblks-1$. However, for simplicity, the CTCP receiver only informs the sender the number of dofs received for block $currblk$.

%%%%%%%%%%%%%%%%%%%%%%%%
\section{CTCP Receiver}\label{sec:receiver}

The receiver is responsible for decoding the received data and providing ACK feedback to the sender.  Whenever the receiver receives a packet, it needs to check whether the current block is decodable ($ack\_currblk$) and to calculate how many dofs it has received for the current block ($ack\_currdof$).

%%%%%%%%%%%%%%%%%%%%%%%%
\subsection{Decoding Operations}\label{sec:decoding}

For each block $blkno$, the receiver initializes a $blksize\times blksize$ matrix $C_{blkno}$ for the coding coefficients and a corresponding payload structure $P_{blkno}$. Whenever a packet from $blkno$ is received, the coding coefficients and the coded payload are inserted to $C_{blkno}$ and $P_{blkno}$ respectively.   Gaussian elimination is then used to determine whether the received packet is linearly independent with the previously received packets.  If it is,  the receiver sets $ack\_currdof \leftarrow ack\_currdof+1$. If $ack\_currdof$ is equal to $blksize$, then the receiver acknowledges that enough dofs have been received for $ack\_currblk$ and updates $ack\_currblk \leftarrow ack\_currblk+1$ ($ack\_currdof$ is also reset to reflect the dofs needed for the new $ack\_currblk$).   If the received packet is not linearly independent from previously received packets, the receiver transmits an ACK (corresponding to the packet received) but does not update $ack\_currdof$ nor $ack\_currblk$.
Once enough $blksizef$ linearly independent packets (dofs)  have been received for a block, the receiver can decode all packets within the block.

%%%%%%%%%%%%%%%%%%%%%%%%
\section{Experimental Measurements}\label{sec:experimental}

In this section, we evaluate CTCP's performance in a testbed. We present results on not only throughput but also on friendliness and fairness.

\begin{figure}
\centering
\includegraphics[width=.43\textwidth]{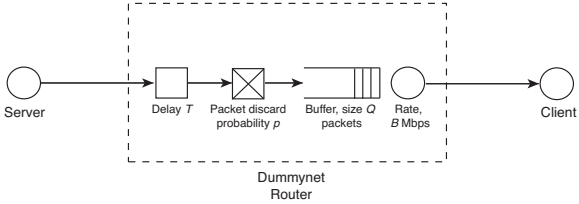}
\vspace*{-.3cm}
\caption{Schematic of experimental testbed.}\label{fig:testbed}\vspace*{-.3cm}
\end{figure}

%%%%%%%%%%%%%%%%%%%%%%%%
\subsection{Testbed Setup}\label{sec:testbed}

\begin{figure}
\centering
\subfloat[Link 25Mbps, RTT 20ms]{
\includegraphics[width=\columnwidth]{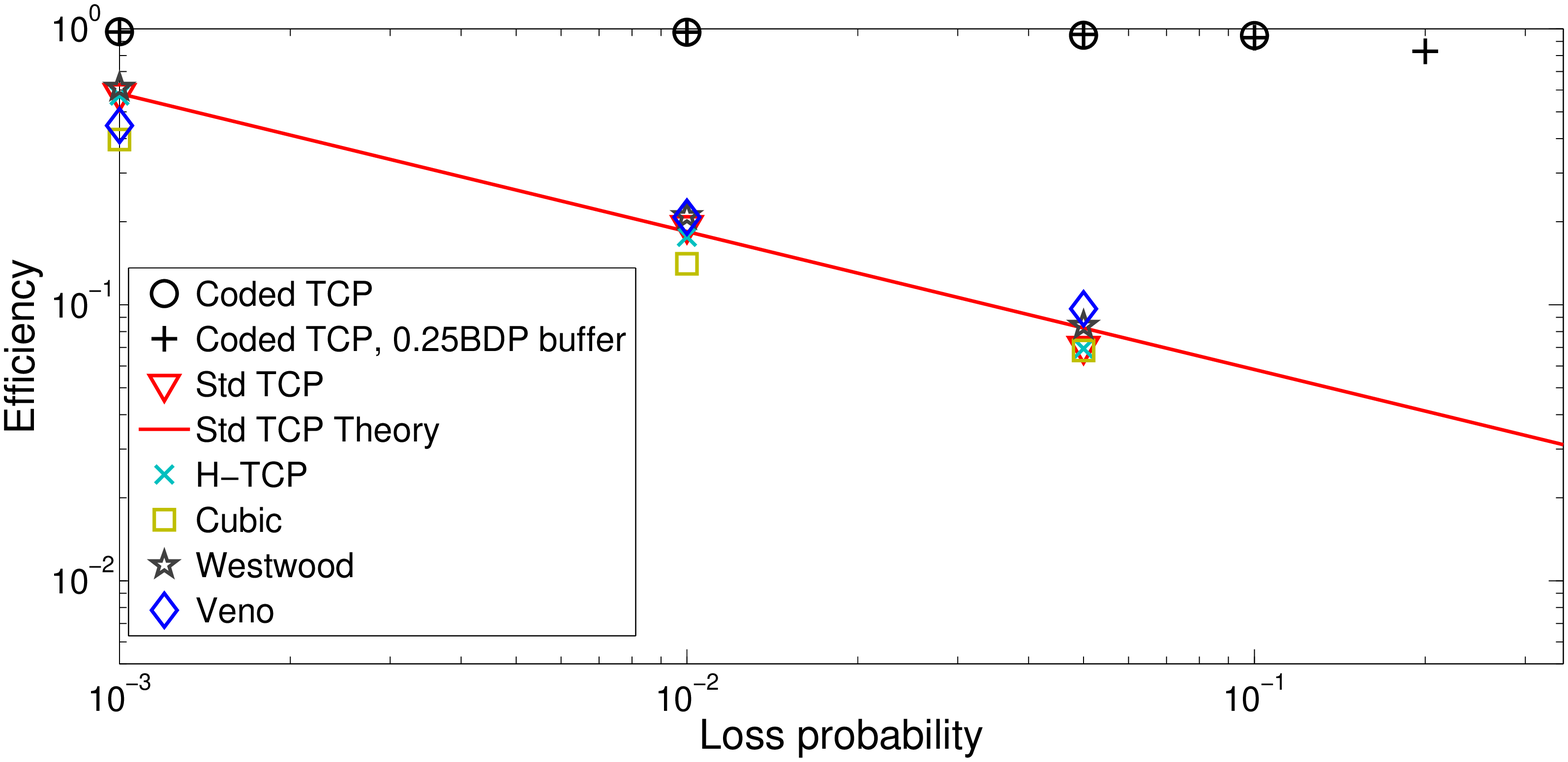}\label{fig:efficiency_A}
}\\
\vspace*{-.1cm}
\subfloat[CTCP]{
\includegraphics[width=\columnwidth]{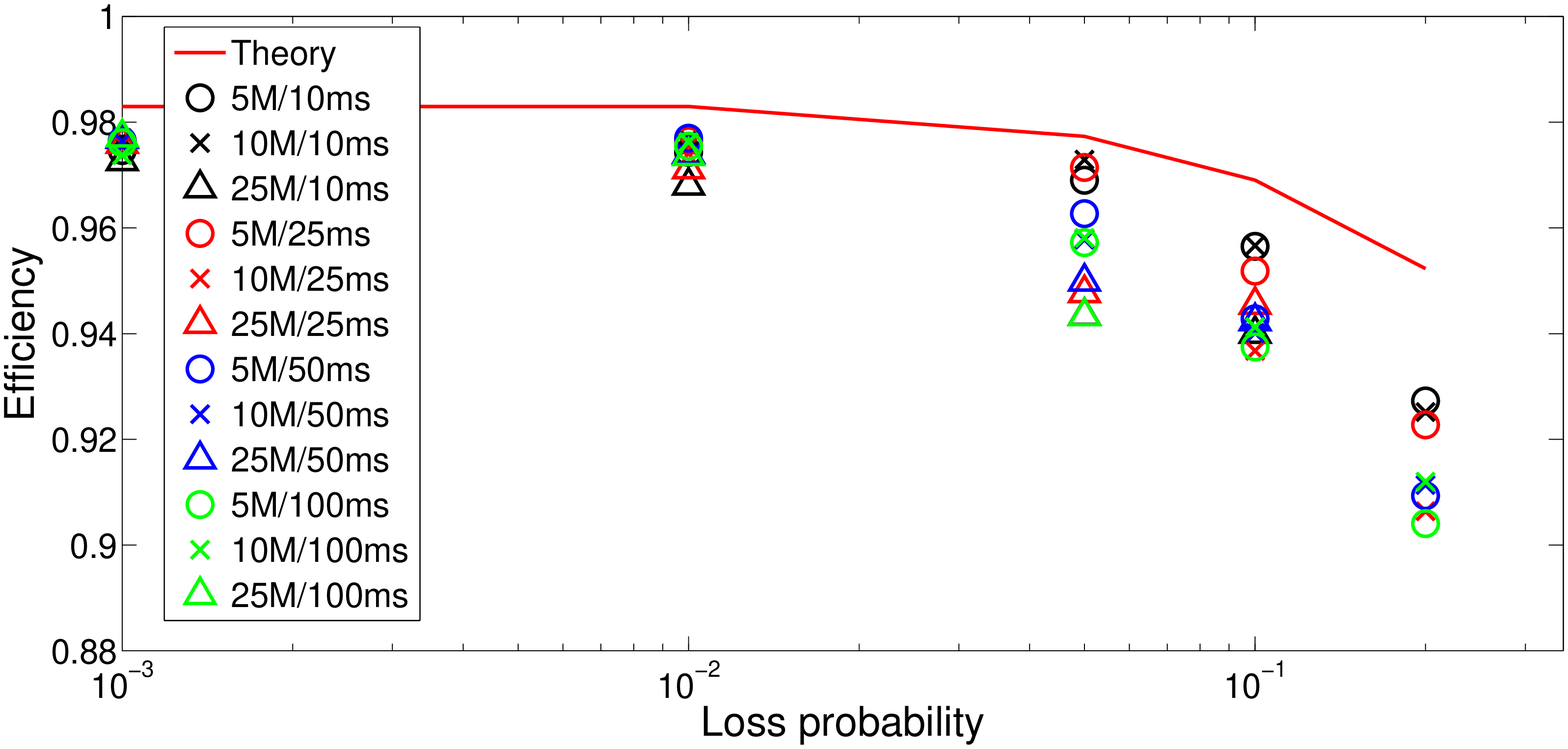}\label{fig:efficiency_B}
}
\caption{Measurements of goodput efficiency against packet loss rate, link rate and RTT. The Theory curve in Figure \ref{fig:efficiency_B} is generated using Equation \eqref{eq:bound}.}\label{fig:efficiency}\vspace*{-.2cm}
\end{figure}

The lab testbed consists of commodity servers (Dell Poweredge 850, 3GHz Xeon, Intel 82571EB Gigabit NIC) connected via a router and gigabit switches (Figure \ref{fig:testbed}).  Sender and receiver machines used in the tests both run a Linux 2.6.32.27 kernel.  The router is also a commodity server running FreeBSD 4.11 and \texttt{ipfw}-\texttt{dummynet}.  It can be configured with various propagation delays $T$, packet loss rates $p$,  queue sizes $Q$ and link rates $B$ to emulate a range of network conditions.  As indicated in Figure \ref{fig:testbed}, packet losses in \texttt{dummynet}  occur before the rate constraint, not after, and so do not reduce the bottleneck link capacity $B$.   Unless otherwise stated, appropriate byte counting is enabled for standard TCP and experiments are run for at least 300s.   Data traffic is generated using \texttt{rsync} (version 3.0.4), HTTP traffic using \texttt{apache2} (version 2.2.8) and \texttt{wget} (version 1.10.2), video traffic using \texttt{vlc} as both server and client (version 0.8.6e as server, version 2.0.4 as client).

CTCP is implemented in userspace as a forward proxy located on the client and a reverse proxy located on the server.   This has the advantage of portability and of requiring neither root-level access nor kernel changes.   Traffic between the proxies is sent using CTCP.   With this setup, a client request is first directed to the local forward proxy. This transmits the request to the reverse proxy, which then sends the request to the appropriate port on the server.   The server response follows the reverse process.   The proxies support the SOCKS protocol and standard tools allow traffic to be transparently redirected via the proxies. In our tests, we used \texttt{proxychains} (version 3.1) for this purpose.

%%%%%%%%%%%%%%%%%%%%%%%%
\subsection{Efficiency}

\begin{figure}
\centering
\includegraphics[width=\columnwidth]{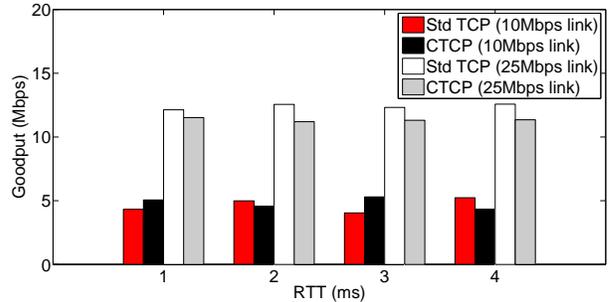}\vspace*{-.3cm}
\caption{Goodput for a standard TCP and a CTCP flow sharing a loss-free link; results are shown for 10Mbps and 25Mbps links with varying RTTs.\vspace{-.3cm}}\label{fig:friendliness}
\end{figure}

\begin{figure}
\centering
\subfloat[Lossy 10Mbps link with RTT=25ms]{\includegraphics[width=0.48\columnwidth]{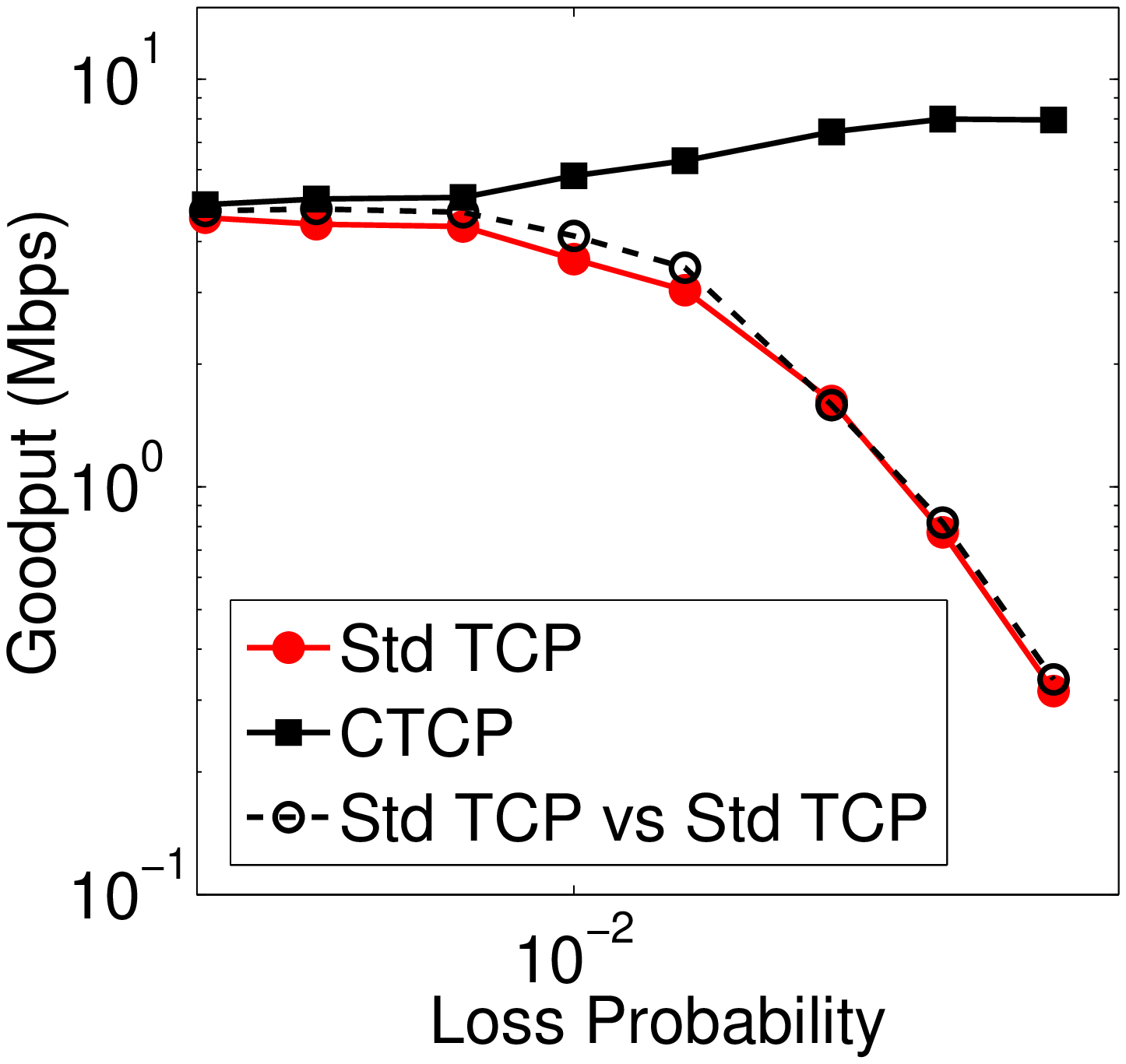}}\hspace*{.2cm}
\subfloat[Lossy 25Mbps link with RTT=25ms]{\includegraphics[width=0.48\columnwidth]{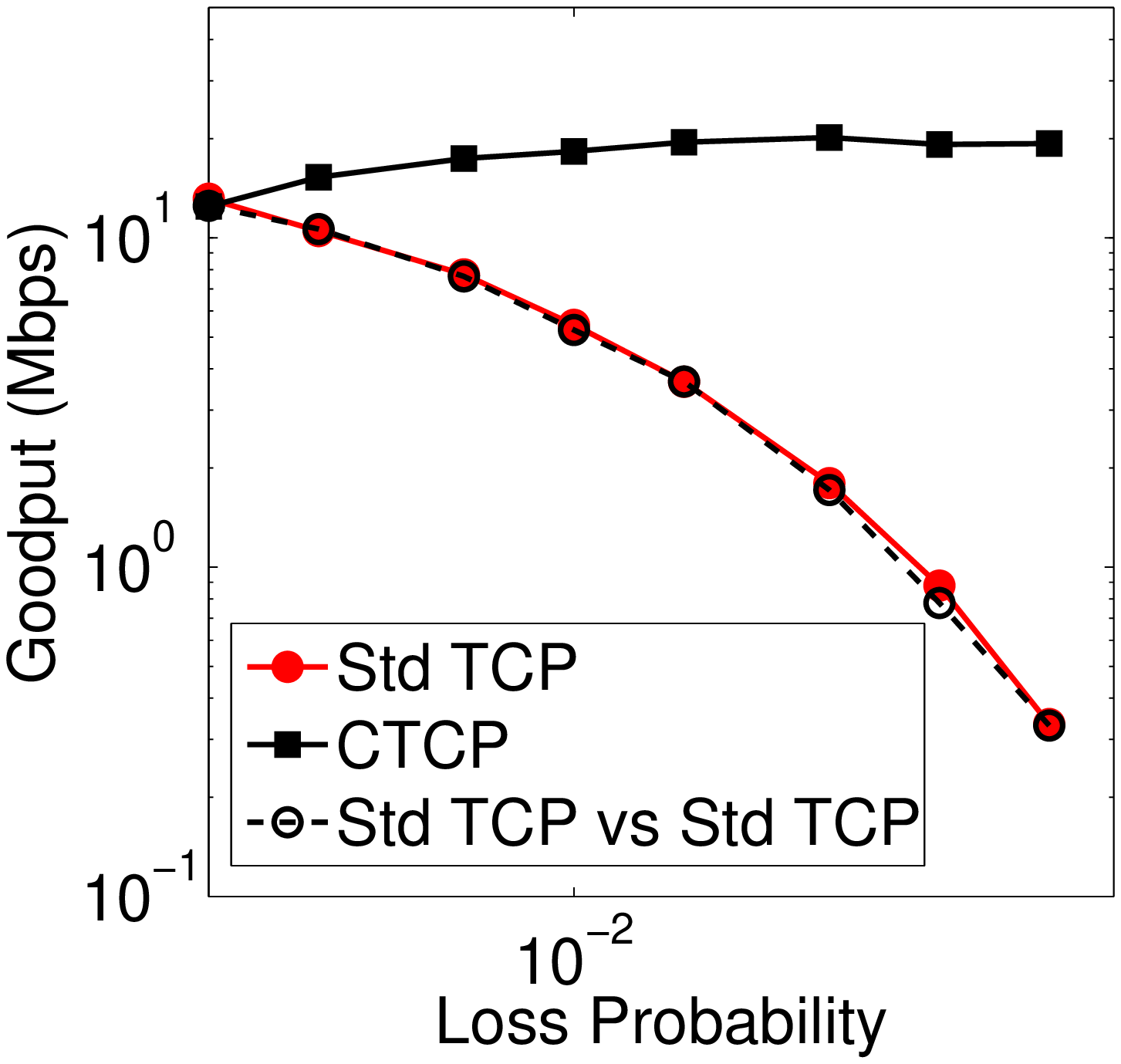}}
\caption{Goodput against link loss rate for (i)   a TCP and  a CTCP flow sharing this link (solid lines), and (ii) two TCP flows sharing lossy link (dashed line).
\vspace{-.3cm}}\label{fig:friendliness2}
\end{figure}

Figure \ref{fig:efficiency} presents experimental measurements of the efficiency (equal to $\frac{\text{goodput}}{\text{link capacity}}$) of various TCP implementations and CTCP over a range of network conditions.   Figure \ref{fig:efficiency_A} shows the measured efficiency versus the packet loss probability $p$ for a 25Mbps link with 25ms RTT and a bandwidth-delay product of buffering.  Baseline data is shown for standard TCP (i.e. TCP SACK/Reno), Cubic TCP (current default on most Linux distributions), H-TCP, TCP Westwood, TCP Veno, together with the value $\sqrt{1.5/p}$ packets per RTT predicted by the popular Padhye model \cite{padhye}.  It can be seen that the measurements for standard TCP are in good agreement with the Padhye model, as expected.  Also that Cubic TCP, H-TCP, TCP Westwood, and TCP Veno closely follow the standard TCP behavior.  Again, this is as expected since the link bandwidth-delay product of 52 packets lies in the regime where these TCP variants seek to ensure backward compatibility with standard TCP.   Observe that the achieved goodput decreases rapidly with increasing loss rate, falling to 20\% of the link capacity when the packet loss rate is 1\%.   This feature of standard TCP is, of course, well known.   Compare this, however, with the efficiency measurements for CTCP, which are shown in Figure \ref{fig:efficiency_A} and also given in  more detail in Figure \ref{fig:efficiency_B}.  The goodput is $>96$\% of link capacity for a loss rate of 1\%, a roughly five-fold increase in goodput compared to standard TCP.

Figure \ref{fig:efficiency_B} presents the efficiency of CTCP for a range of link rates, RTTs and loss rates.   It shows that the efficiency achieved is not sensitive to the link rate or RTT.   Also shown in Figure \ref{fig:efficiency_B} is a theoretical upper bound on the efficiency calculated using
\begin{align}
\eta = \frac{1}{N}\sum_{k=0}^{n-1} (n-k) {n\choose k}  p^k(1-p)^{N-k},\label{eq:bound}
\end{align}
where $N=32$ is the block size, $p$ the packet erasure probability and $n=\lfloor N/(1-p)\rfloor - N$ is the number of forward-transmitted coded packets sent with each block.     This value $\eta$ is the mean number of such forward-transmitted coded packets that are unnecessary (because there are fewer then $n$ erasures).

The efficiency achieved by CTCP is also insensitive to the buffer provisioning, as discussed in Section \ref{sec:congestion}.    This property is illustrated in Figure \ref{fig:efficiency_A}, which presents CTCP measurements when the link buffer is reduced in size to 25\% of the bandwidth-delay product.  The efficiency achieved with 25\% buffering is close to that with a full bandwidth-delay product of buffering.

%%%%%%%%%%%%%%%%%%%%%%%%
\subsection{Friendliness with Standard TCP}\label{sec:friendliness}

\begin{figure}
\centering
\subfloat[25Mbps, RTT 25ms, 5\% packet loss rate]{
\includegraphics[width=0.98\columnwidth]{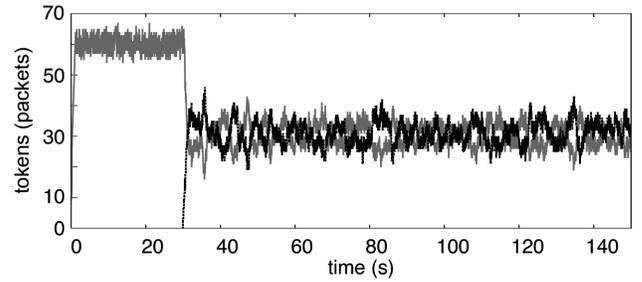}\label{fig:fairness_A}
}\\
\subfloat[ RTT 25ms]{
\includegraphics[width=\columnwidth]{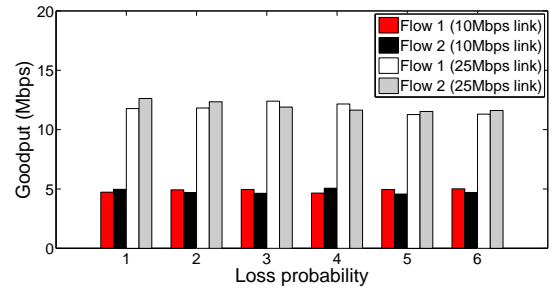}\label{fig:fairness_B}
}\vspace*{-.3cm}
\caption{Measurements of $tokens$ and goodput for two CTCP flows sharing a lossy link.  Figure \ref{fig:fairness_A} provides a sample of $tokens$ time history while Figure \ref{fig:fairness_B} summarizes the goodputs for a range of packet loss rates (\%) with 10Mbps and 25Mbps links (RTT 25ms). Similar behavior was observed for other values of RTTs.\vspace{-.3cm}} \label{fig:fairness}
\end{figure}

Figures \ref{fig:friendliness} and \ref{fig:friendliness2} confirm that standard TCP and CTCP can coexist in a well-behaved manner.   In these measurements, a standard TCP flow and a CTCP flow share the same link competing for bandwidth.   As a baseline, Figure \ref{fig:friendliness} presents the goodputs of TCP and CTCP for range of RTTs and link rates on a \emph{loss-free} link (i.e. when queue overflow is the only source of packet loss).   As expected, it can be seen that the standard TCP and CTCP flows consistently obtain similar goodput.

Figure \ref{fig:friendliness2} presents goodput data when the link is lossy.   The solid lines indicate the goodputs achieved by the CTCP flow and the standard TCP flow sharing the same link with varying packet loss rates.   At low loss rates, they obtain similar goodputs; but as the loss rate increases, the goodput of standard TCP rapidly decreases (as already observed in Figure \ref{fig:efficiency_A}).

For comparison, in Figure \ref{fig:friendliness2}, we also show (using the dotted lines) the goodput achieved by a standard TCP flow when competing against another standard TCP flow (i.e. when two standard TCP flows share the link). Note that the goodput achieved by a standard TCP flow (dotted line) when competing against another standard TCP flow is close to that achieved when sharing the link with a CTCP flow (solid line). This demonstrates that CTCP does not penalize the standard TCP flow.

%%%%%%%%%%%%%%%%%%%%%%%%
\subsection{Fairness among CTCP Flows}

We turn now to fairness, i.e. how goodput is allocated between competing CTCP flows.   Figure \ref{fig:fairness_A} plots a typical $tokens$ time history for two CTCP flows sharing a lossy link. The second flow (black) is started after the first (grey) so that we can observe the convergence to fairness.   It can be seen that the two flows' $tokens$ rapidly converge. Figure \ref{fig:fairness_B} presents corresponding goodput measurements for a range of link rates, RTTs, and loss rates.   Again, the two CTCP flows consistently achieve similar goodputs.

%%%%%%%%%%%%%%%%%%%%%%%%
\subsection{Application Performance}

In this section, we present our testbed results seen by various applications.

%%%%%%%%%%%%%%%%%%%%%%%%
\subsubsection{Web}

\begin{figure}
\centering
\includegraphics[width=\columnwidth]{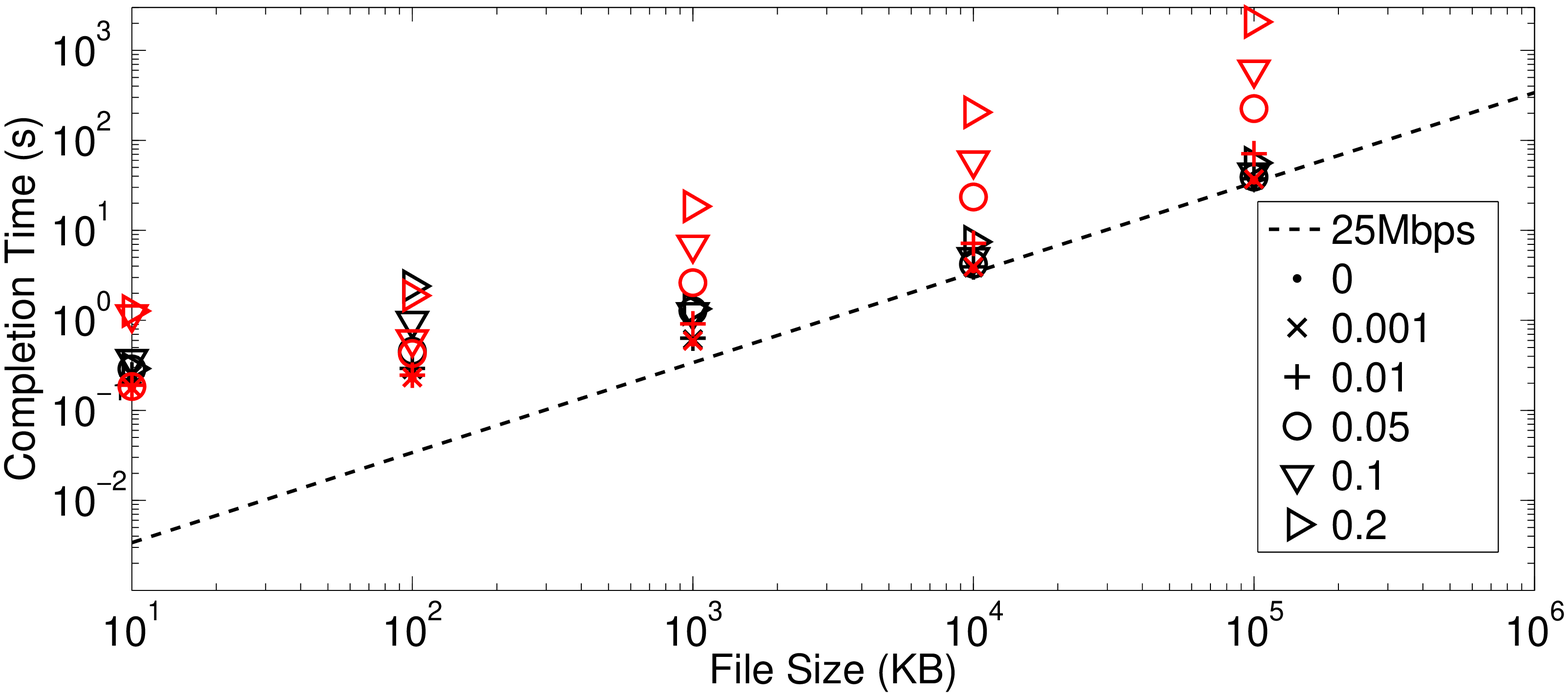}\vspace*{-.3cm}
\caption{Measured HTTP request mean completion time against file size over 25Mbps link with RTT = 10ms.   Data is shown for standard TCP (red) and CTCP (black) for a range of loss rates.  Error bars are comparable in size to the symbols used in the plot and so are omitted.}\label{fig:http}\vspace*{-.3cm}
\end{figure}

Figure \ref{fig:http} shows measurements of HTTP request completion time against file size for standard TCP and CTCP. The HTTP requests are generated using \texttt{wget} and the response is by an \texttt{apache2} web server. Note the log scale on the y-axis.

The completion times with CTCP are largely insensitive to the packet loss rate. For larger file sizes, the completion times approach the best possible performance indicated by the dashed line. For smaller file sizes, the completion time is dominated by slow-start behavior.   Note that CTCP and TCP achieve similar performance when the link is loss-free; however, TCP's completion time quickly increases with loss rate.   For a 1MB connection, the completion time with standard TCP increases from 0.9s to 18.5s as the loss rate increases from 1\% to 20\%, while for a 10MB connection the corresponding increase is from 7.1s to 205s.  Observe that the completion time is reduced by more than 20$\times$ (2000\%) for a 1MB connection and by almost $30\times$ (3000\%) for a 10MB connection.

%%%%%%%%%%%%%%%%%%%%%%%%
\subsubsection{Streaming Video}
\begin{figure}
\vspace*{-.3cm}
\centering
\subfloat[Completion Time]{
\includegraphics[width=0.48\columnwidth]{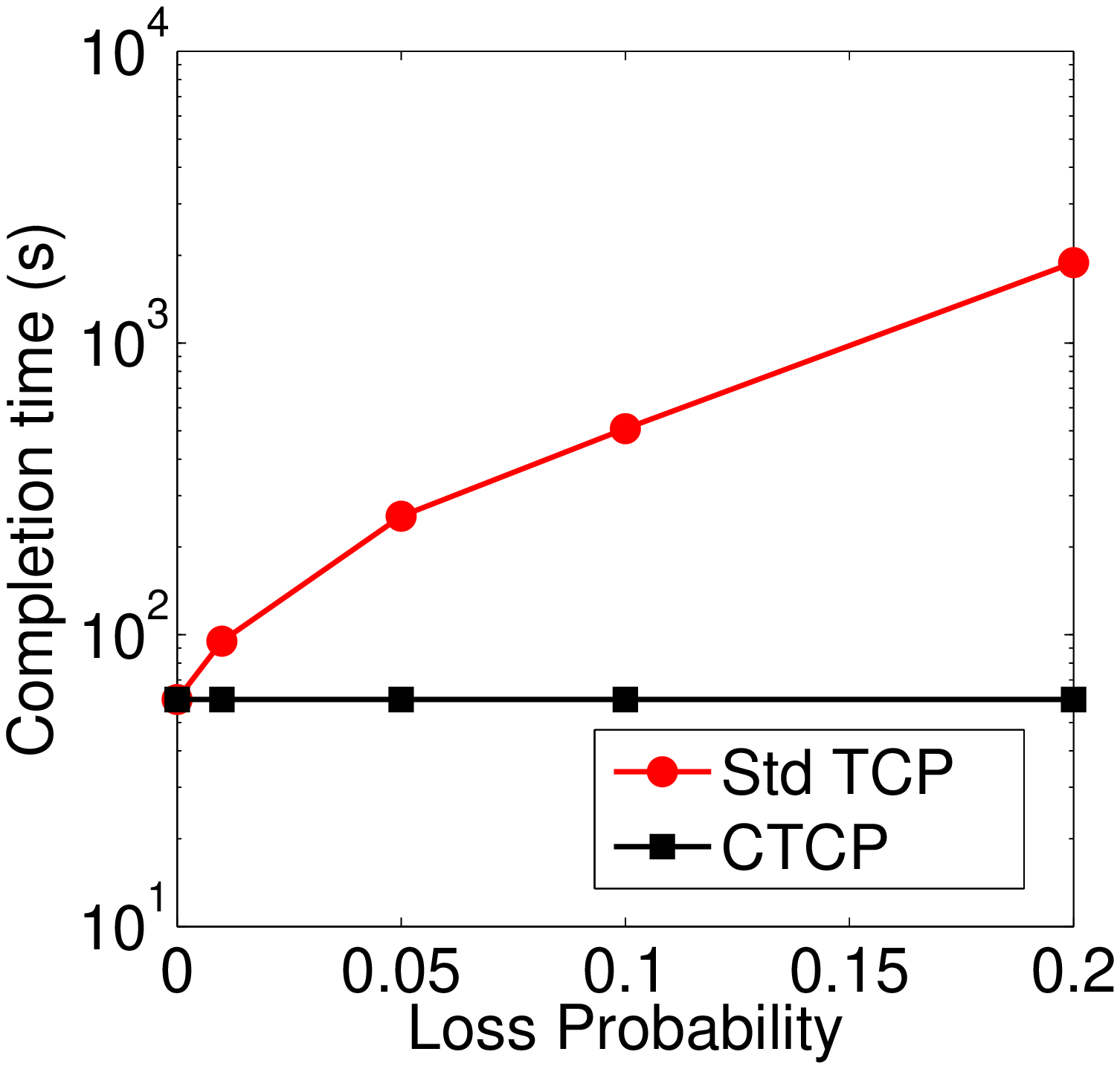}\label{fig:video_A}
}
\subfloat[Buffer Under-runs]{
\includegraphics[width=0.48\columnwidth]{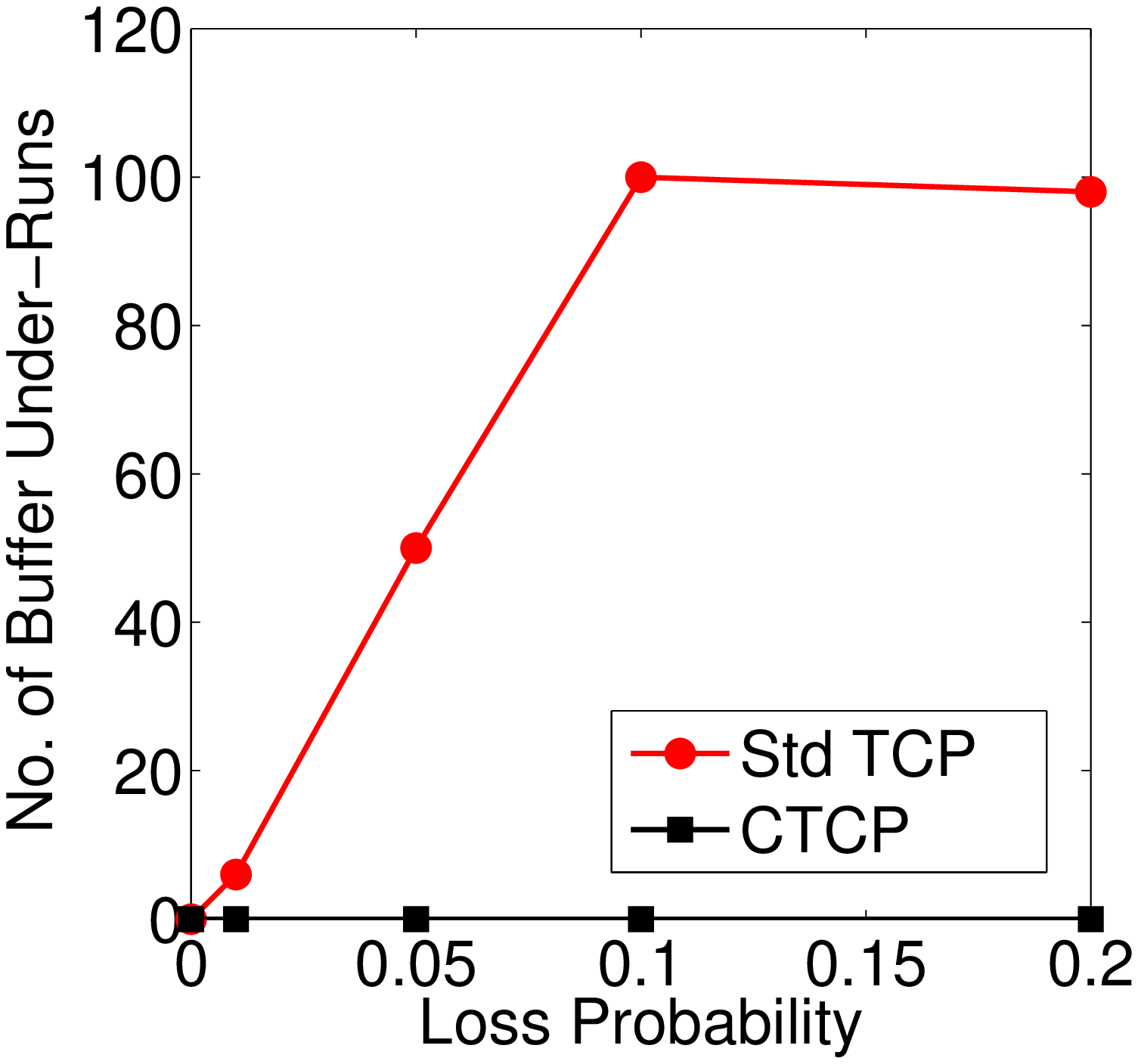}\label{fig:video_B}
}
\caption{Measurements of video streaming performance against loss rate with a 25Mbps link and a RTT of 10ms.   Data is shown for standard TCP and CTCP.  Figure \ref{fig:video_A} shows the running time taken to play a video of nominal duration (60s); Figure \ref{fig:video_B} shows the number of under-runs of the playout buffer at the client.}\label{fig:video}\vspace*{-.3cm}
\end{figure}

Figure \ref{fig:video} plots performance measurements for streaming video for a range of packet loss rates on a 25Mbps link with RTT equal to 10 ms.   A \texttt{vlc} server and client are used to stream a 60s video.  Figure \ref{fig:video_A} plots the measured time for playout of the video to complete.  Again, note the log scale on the y-axis.

The playout time with CTCP is approximately 60s and is insensitive to the packet loss rate.   In contrast, the playout time with standard TCP increases from 60s to 95s when the loss rate is increased from 0\% to 1\%, and increases further to 1886s (31 minutes) as the loss rate is increased to 20\% (more than 30$\times$ slower than when using CTCP).   Figure \ref{fig:video_B} plots measurements of playout buffer under-run events at the video client.  It can be seen that there are no buffer under-run events when using CTCP even when the loss rate is as high as 20\%.   With standard TCP, the number of buffer under-runs increases with loss rate until it reaches a plateau at around 100 events, corresponding to a buffer underrun occurring after every playout of a block of frames.  In terms of user experience, the increases in running time result in the video repeatedly stalling for long periods of time and are indicative of a thoroughly unsatisfactory quality of experience even at a loss rate of 1\%.

%%%%%%%%%%%%%%%%%%%%%%%%
\section{Real-world Performance}\label{sec:realworld}

In this section we present measurements from a number of wireless links subject to real impairments.

\subsection{Microwave Oven Interference}\label{sec:mwo}

We begin by considering an 802.11b/g wireless client downloading from an access point over a link subject to interference from a domestic microwave oven (MWO).  The wireless client and AP were equipped with Atheros 802.11 b/g 5212 chipsets (radio 4.6, MAC 5.9, PHY 4.3 using Linux MadWifi driver version 0.9.4) operating on channel 8.  The microwave oven used was a 700 W Tesco MM08 17L.   Unless otherwise stated, the default operating system settings are used for all network parameters.  The wireless client used rsync to download a 50MB file via the AP.   Figure \ref{fig:mwo1} (shown in the introduction) presents spectrum analyzer  (Rohde \& Schwarz FSL-6) measurements illustrating both the MWO interference and packet transmissions on the wireless link.   The MWO operates in the 2.4 GHz ISM band, with significant overlap ($>50\%$) with the WiFi 20 MHz channels 6 to 13.  The MWO interference is approximately periodic, with period 20ms (i.e. 50Hz) and mean pulse width 9ms (the width was observed to fluctuate due to frequency instability of the MWO cavity magnetron, a known effect in MWOs).

\begin{figure}
\centering
\includegraphics[width=\columnwidth]{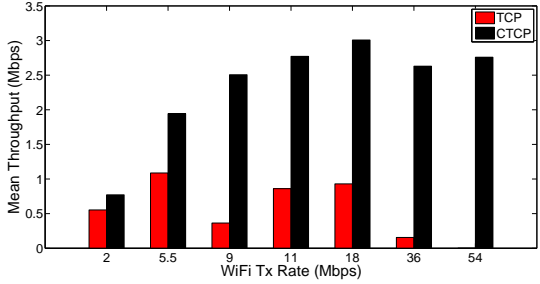}
\caption{Measurements of mean throughput vs wireless PHY rate used with standard TCP (Cubic TCP) and CTCP on an 802.11 link affected by microwave oven interference.\vspace{-.3cm}}\label{fig:mwo2}
\end{figure}

Figure \ref{fig:mwo2} presents measurements of the mean throughout achieved over the file download vs the PHY rate used on the downlink.    Data is shown using standard TCP (in this case Cubic TCP, the Linux default variant) and CTCP.   Data is not shown for a PHY rate of 1Mbps as the packet loss rate was close to 100\% -- this is because at this PHY rate, the time to transmit a 1500B frame is greater than the interval between MWO interference bursts and so almost all frames are damaged by the interference.   It can be seen that the throughput achieved by standard TCP rises slightly as the PHY rate is increased from 1Mbps to 5.5Mbps, but then decreases to zero for PHY rates above 36Mbps (due to channel path losses). In comparison, when using CTCP the throughput achieved is approximately doubled (200\%) at a PHY rate of 5.5Mbps, more than tripled (300\%) at PHY rates of 8, 11 and 18 Mbps and  increased by more than an order of magnitude (1000\%) at a PHY rates above this.  Furthermore, the fluctuations of both TCP and CTCP performance under different link layer coding rates and modulation schemes (indicated by the changes in the WiFi transmission rate) suggests that CTCP's performance is much more robust to network changes than that of TCP's performance, although more testing is required to verify this claim.

\subsection{Hidden Terminal Interference}

\begin{figure}
\centering
\includegraphics[width=\columnwidth]{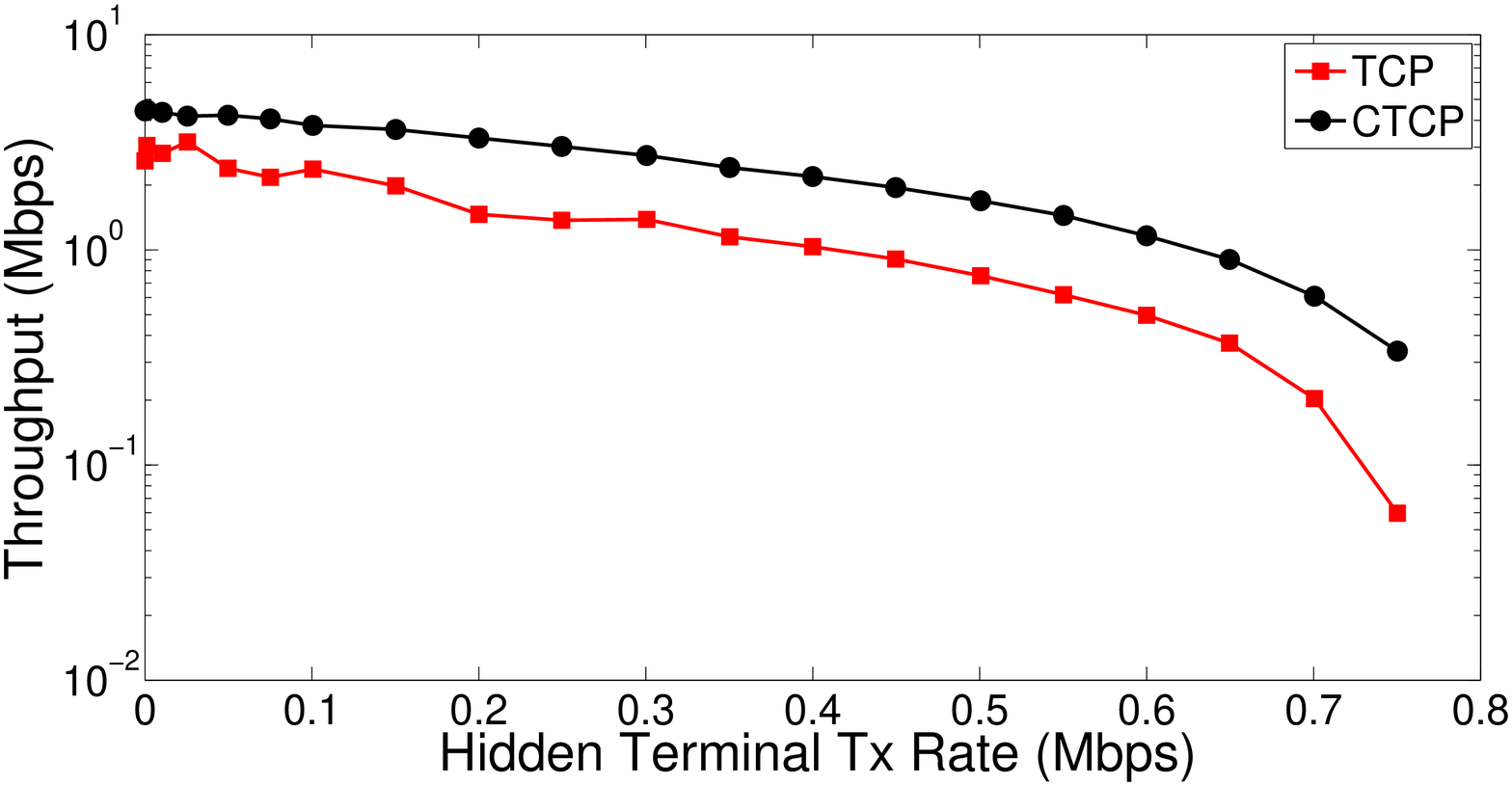}
\caption{Throughput vs intensity of hidden terminal interference when using standard TCP (Cubic TCP) and CTCP over an 802.11b/g wireless link.\vspace{-.3cm}}\label{fig:hidden}
\end{figure}

We now consider an 802.11 wireless link (configured similarly to that in Section \ref{sec:mwo}) which is subject to hidden terminal interference.   The hidden terminal is created by adding a third station to the network created in the last section.  Carrier sense on the terminal's wireless interface card is disabled and 1445B UDP packets are transmitted with exponentially distributed inter-arrival times. The 802.11 transmit rates for both the hidden terminal and AP were set to 11 Mbps.  Unless otherwise stated, the default operating system settings are used for all of the remaining network parameters. Figure \ref{fig:hidden} plots the measured throughput on the downlink from the AP to a wireless client versus the mean transmit rate of the hidden terminal traffic.    It can be seen that CTCP consistently obtains approximately twice (200\%) the throughput of standard TCP (Cubic TCP) across a wide range of interference conditions.   

\subsection{Public WiFi Measurements}

\begin{figure*}
\centering
\subfloat[ \small CTCP Time = 313 s, \newline TCP Time = 807 s, \newline Mean PLR = 4.28\%, \newline Mean RTT = 54.21 ms]{
\includegraphics[scale=0.18]{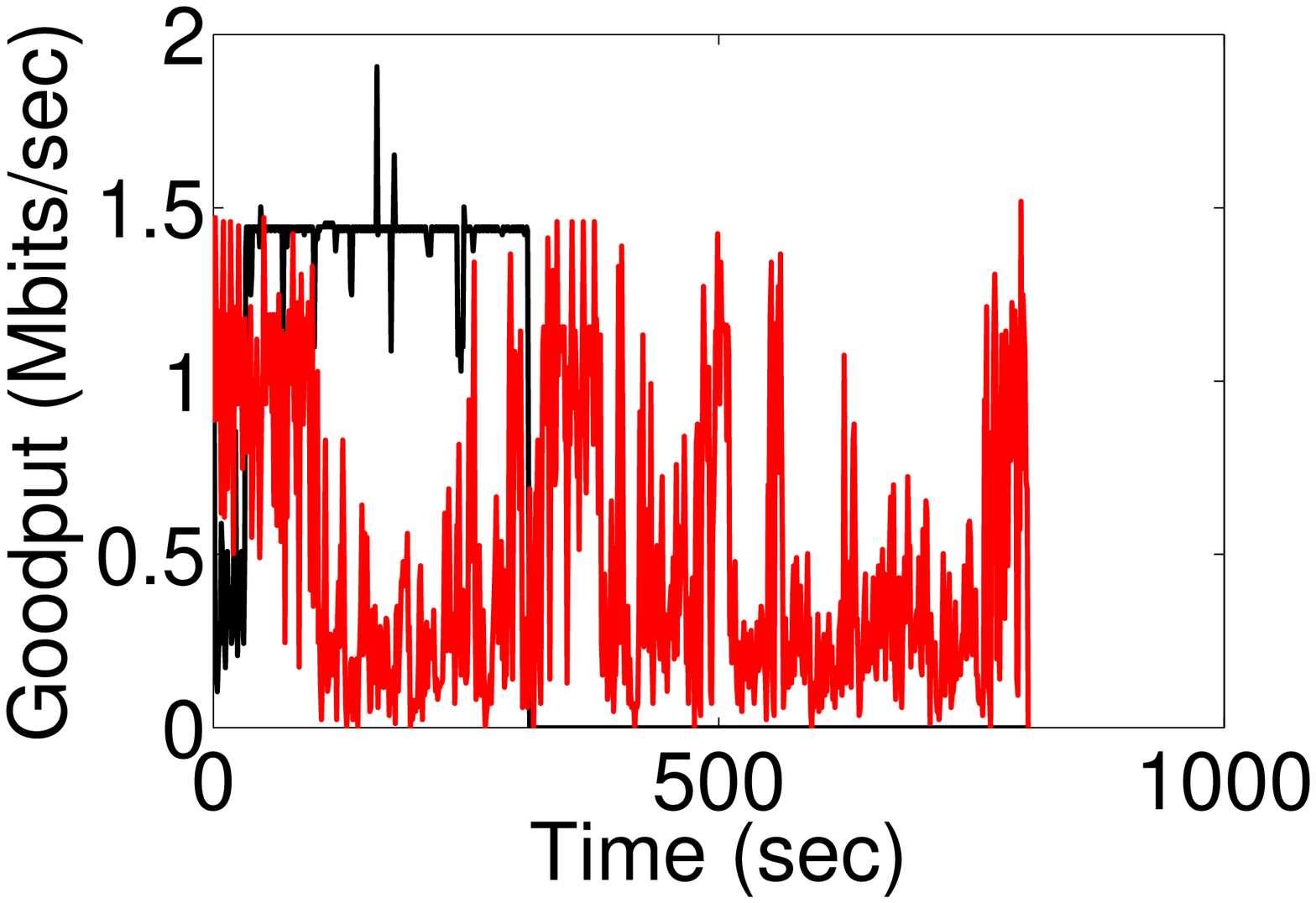}\label{fig:wifi_tp_A}
}
\subfloat[ \small CTCP Time = 388 s, \newline TCP Time = 1151 s, \newline Mean PLR = 5.25\%, \newline Mean RTT = 73.51 ms]{
\includegraphics[scale=0.18]{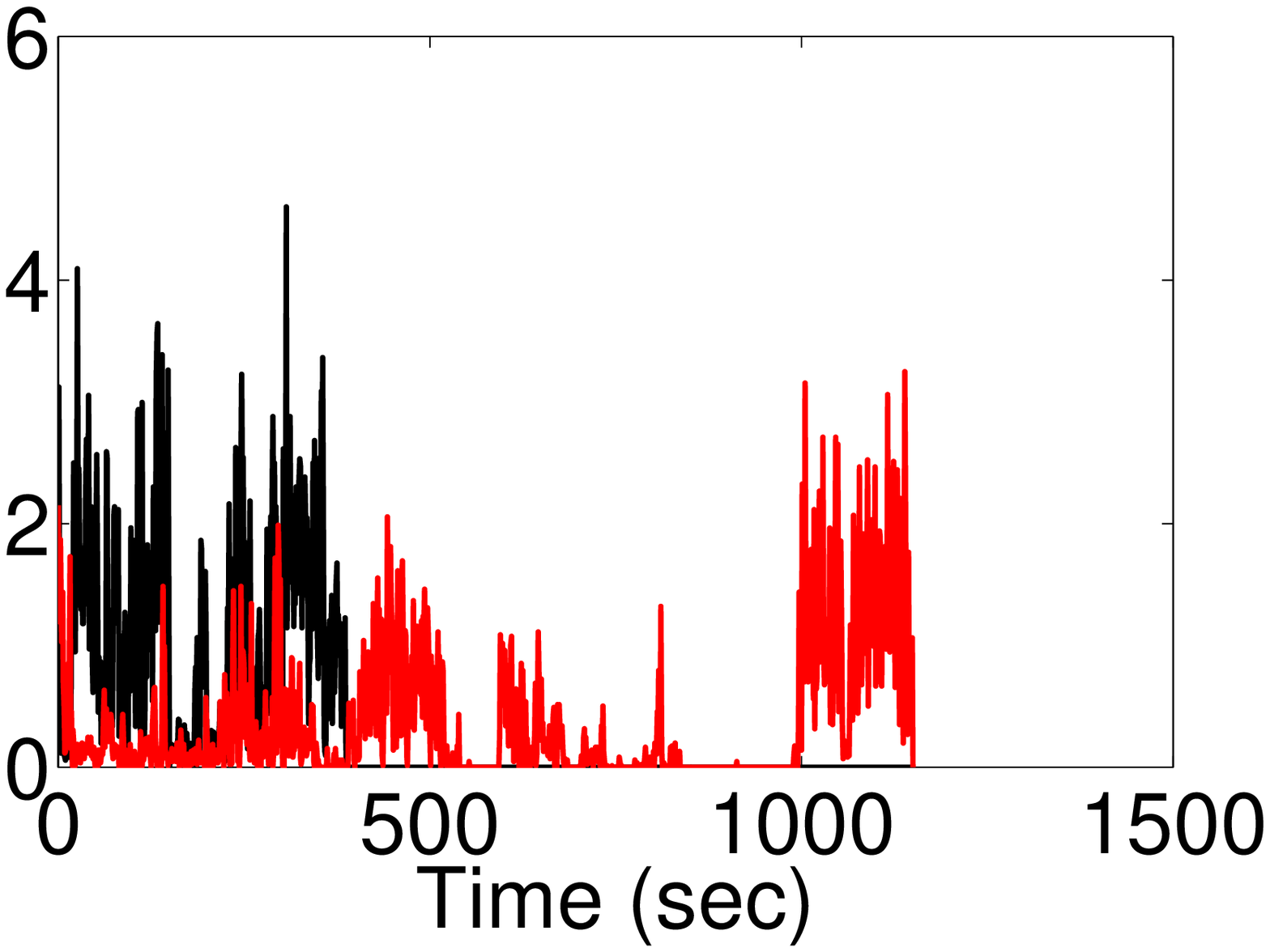}\label{fig:wifi_tp_B}
}
\subfloat[ \small CTCP Time = 676 s, \newline TCP Time = 753 s, \newline Mean PLR = 4.65\%, \newline Mean RTT = 106.31 ms]{
\includegraphics[scale=0.18]{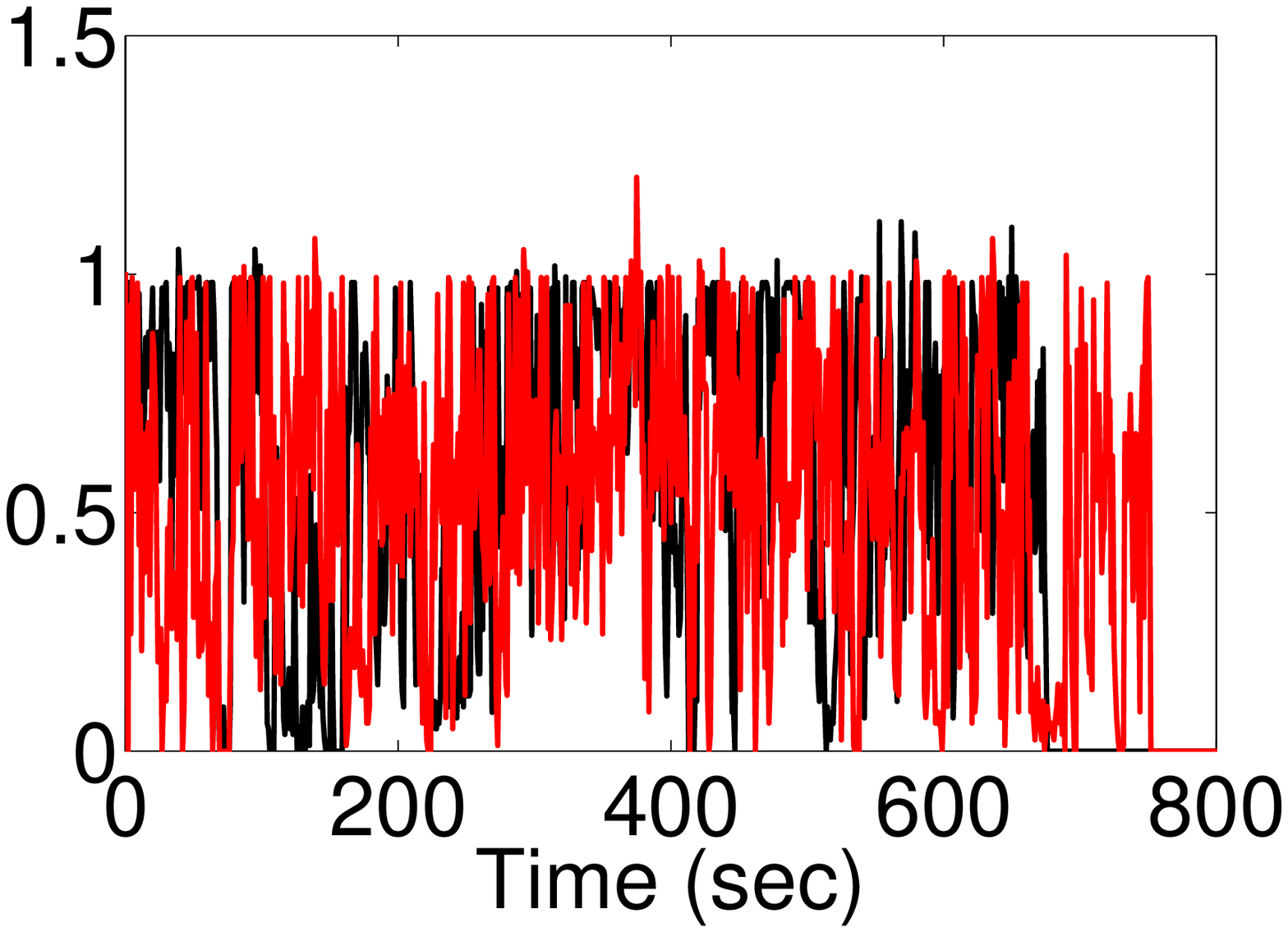}\label{fig:wifi_tp_C}
}
\subfloat[ \small CTCP Time = 292 s, \newline TCP Time = 391 s, \newline Mean PLR = 4.56\%, \newline Mean RTT = 50.39 ms]{
\includegraphics[scale=0.18]{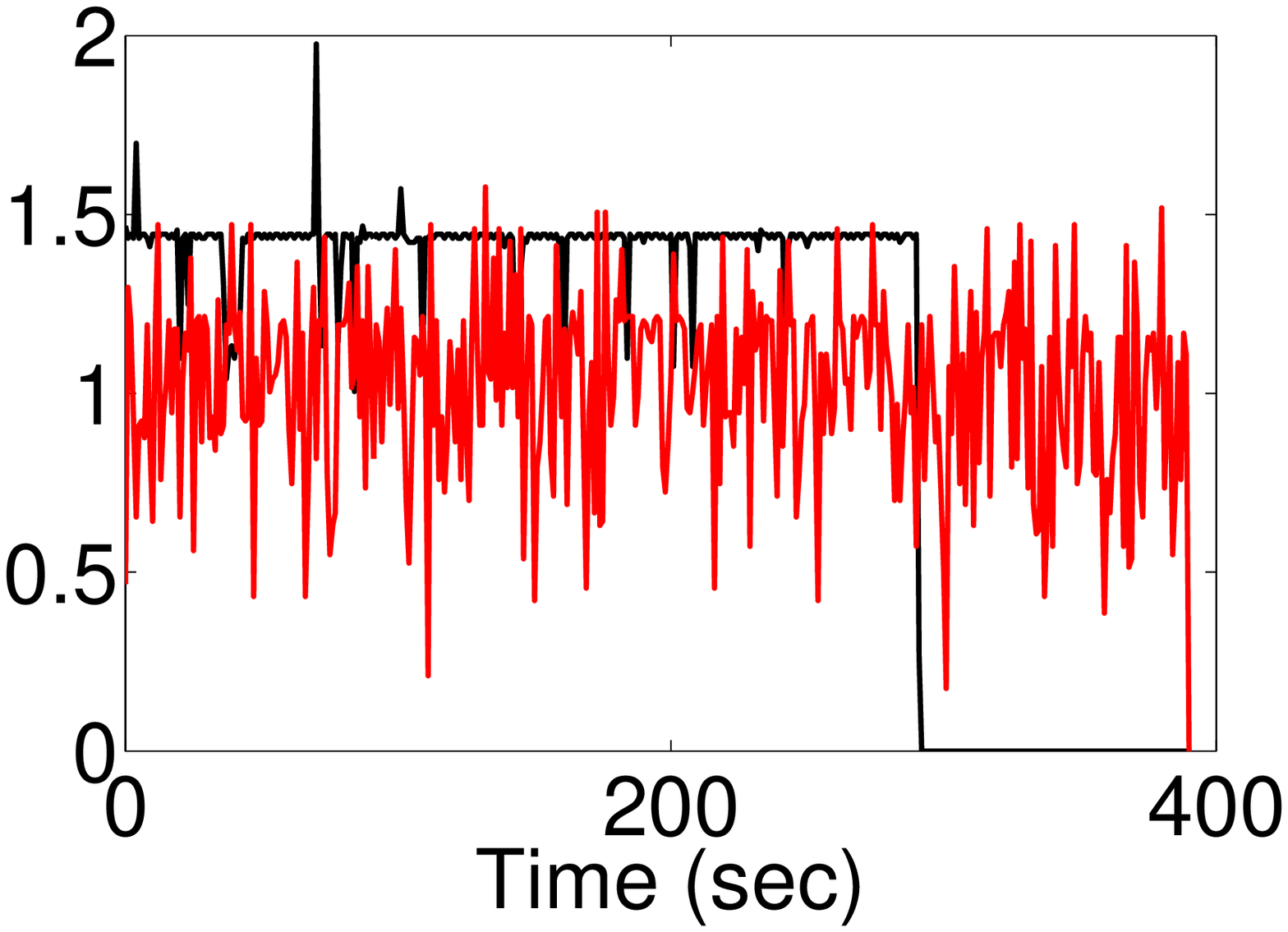}\label{fig:wifi_tp_D}
}
\subfloat[ \small CTCP Time = 1093 s, \newline TCP Time = 3042 s, \newline Mean PLR = 2.16\%, \newline Mean RTT = 208.94 ms]{
\includegraphics[scale=0.18]{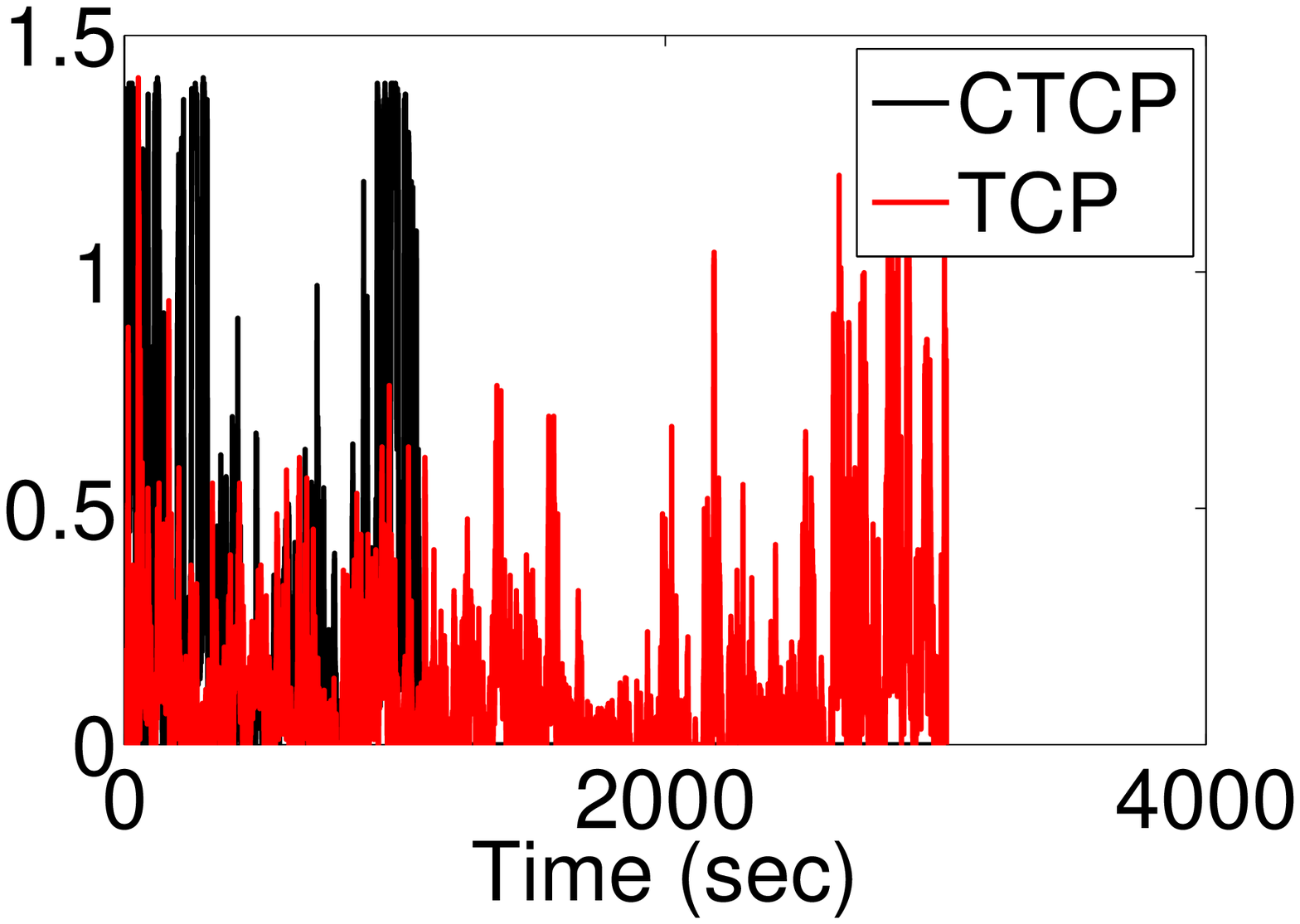}\label{fig:wifi_tp_E}
}
\caption{Public WiFi Network Test Traces (CTCP in black, TCP in red). The download completion times, the mean packet loss rate ($PLR$), and mean $RTT$ for each experiment are also provided. \label{fig:wifi_tp}}
\vspace{-.3cm}
\end{figure*}

Measurements were collected at various public WiFi networks in the greater Boston area by downloading a 50 MB file from a server (running Ubuntu 10.04.3 LTS) located on the MIT campus to a laptop (running Ubuntu 12.04.1 LTS) under the public WiFi hotspot.     The default operating system settings are used for all network parameters on client and server. 
Figure \ref{fig:wifi_tp} shows representative traces for five examples of these experiments. It is important to point out that standard TCP stalled and had to be restarted twice before successfully completing in the test shown in Figure \ref{fig:wifi_tp_C}.  CTCP, on the other hand, never stalled nor required a restart.

Each trace represents a different WiFi network that was chosen because of the location, accessibility, and perceived congestion.  For example, the experiments were run over WiFi networks in shopping center food courts, coffee shops, and hotel lobbies.  In Figures \ref{fig:wifi_tp_A} - \ref{fig:wifi_tp_D}, the WiFi network spanned a large user area increasing the possibility of hidden terminals; a scan of most of the networks showed > 40 active WiFi radios, which also increases the probability of collision losses. The only experiment that had a small number of terminals (i.e. five active radios) is shown in Figure \ref{fig:wifi_tp_E}.   The mean packet loss rate measured over all experiments was approximately $4\%$.

\begin{figure}
\includegraphics[width=\columnwidth]{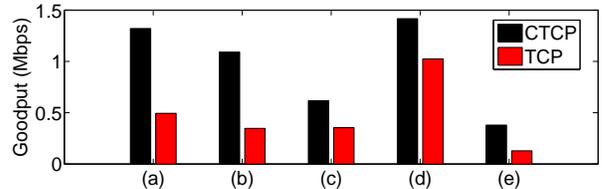}
\caption{Mean goodput for each of the experiments shown in Figure \ref{fig:wifi_tp}.\label{fig:wifi_avg_tp_t}}
\vspace{-.3cm}
\end{figure}

It can be seen that in each of the experiments, CTCP consistently achieved a larger average goodput and faster completion time.  The average throughput for both CTCP and TCP is shown in Figure \ref{fig:wifi_avg_tp_t}. Taking the mean throughput over all of the conducted experiments, CTCP achieves a goodput of approximately $750$ kbps while standard TCP achieves approximately $300$ kbps; resulting in a gain of approximately $2.5$ (250\%).  

We emphasize the observed loss rates of approximately $4\%$ in Figure \ref{fig:wifi_tp}, which is quite high and unexpected; resulting in CTCP's significant performance gain over TCP's. We believe that the loss rate is not only due to randomness but also due to congestion, interference, and hidden terminals. This is an area that would be worthwhile to investigate further. If our intuition is indeed correct, we believe that CTCP can greatly help increase efficiency in challenged network environments.

\section{Summary and Future Work}\label{sec:scope}

We introduce CTCP, a reliable transport protocol using network coding.  CTCP is designed to incorporate TCP features such as congestion control and reliability while significantly improving on TCP's performance in lossy, interference-limited and/or dynamic networks. A key advantage of adopting a transport layer over a link layer approach is that it provides backward compatibility with the enormous installed base of existing wireless equipment.   We present an portable userspace implementation of CTCP (requiring neither kernel level modifications nor changes to the network) and extensively evaluate its performance in both testbed and production wireless networks.   In controlled lab experiments, we consistently observed reductions of more than an order of magnitude (i.e. >1000\%) in completion times for both HTTP and streaming video flows when the link packet loss rate exceeds 5\%.   Measurements in public WiFi hotspots  demonstrate that CTCP can achieve reductions in connection completion times of 100-300\% compared with uncoded TCP.   

There are areas for further research. CTCP's congestion control mechanism performs well in lossy networks where an increase in RTT indicates congestion.  When the RTT also varies with non-congestion events such as variability resulting from specific MAC implementations, CTCP's congestion control can needlessly limit throughput.  New approaches for fairness and friendliness may be needed for such networks.  In addition, we did not investigate the potential impact of active queue management (AQM) on CTCP.  However, the effect of AQM may not be significant as fewer networks use AQMs with the introduction of protocols using streamlets, selective repeat mechanisms, and new congestion control mechanisms such as Cubic. Another possible extension is to allow re-encoding within the network \cite{HMKKESL06,LMKE08,xor}, although this may require changes within the network (not just end-to-end).  However, this approach has been shown to increase efficiency. Finally, we believe that CTCP can be extended to provide gains in multi-path environments \cite{multipathctcp}.  By coding over multiple paths, initial simulations and experiments in \cite{multipathctcp} show that we can achieve the sum rate of each path without the complexity of tracking and scheduling individual packets through the multiple networks.

\bibliography{ctcp_bibV4}
\bibliographystyle{ieeetr}

\end{document}